\documentclass{article} % For LaTeX2e
\usepackage{iclr2021_conference,times}

\usepackage{amsmath,amsfonts,bm}

\def\eqref#1{equation~\ref{#1}}
\def\1{\bm{1}}

\DeclareMathAlphabet{\mathsfit}{\encodingdefault}{\sfdefault}{m}{sl}
\SetMathAlphabet{\mathsfit}{bold}{\encodingdefault}{\sfdefault}{bx}{n}

\usepackage{hyperref}
\usepackage{url}
\usepackage{inconsolata}
\usepackage{color}
\usepackage{enumitem}
\usepackage{graphicx}
\usepackage{float}
\usepackage[section]{placeins}
\usepackage{subcaption}
\usepackage{pdfpages}
\usepackage{chngpage}
\usepackage{amsmath}

\title{Fairness in Image Search: A Study of Occupational Stereotyping in Image Retrieval  and its Debiasing}

\author{Swagatika Dash\\
Information School\\
University of Washington\\
Seattle, WA 98105, USA\\
\texttt{sdash2@uw.edu}
}

\iclrfinalcopy % Uncomment for camera-ready version, but NOT for submission.
\begin{document}

\maketitle

\begin{abstract}
Multi-modal search engines have experienced significant growth and widespread use in recent years, making them the second most common internet use. While search engine systems offer a range of services, the image search field has recently become a focal point in the information retrieval community, as the adage goes, \textit{a picture is worth a thousand words.} Although popular search engines like Google excel at image search accuracy and agility, there is an ongoing debate over whether their search results can be biased in terms of gender, language, demographics, socio-cultural aspects, and stereotypes. This potential for bias can have a significant impact on individuals' perceptions and influence their perspectives.

In this paper, we present our study on bias and fairness in web search, with a focus on keyword-based image search. We first discuss several kinds of biases that exist in search systems and why it is important to mitigate them. We narrow down our study to assessing and mitigating occupational stereotypes in image search, which is a prevalent fairness issue in image retrieval. For the assessment of stereotypes, we take gender as an indicator. We explore various open-source and proprietary APIs for gender identification from images. With these, we examine the extent of gender bias in top-tanked image search results obtained for several occupational keywords. To mitigate the bias, we then propose a fairness-aware re-ranking algorithm that optimizes (a) relevance of the search result with the keyword and (b) fairness \textit{w.r.t} genders identified. We experiment on 100 top-ranked images obtained for 10 occupational keywords and consider random re-ranking and re-ranking based on relevance as baselines. Our experimental results show that the fairness-aware re-ranking algorithm produces rankings with better fairness scores and competitive relevance scores than the baselines.
\end{abstract}

\noindent
\noindent \textbf{Report Organization:} The paper is organized as follows. 
Section \ref{sec:intro} is the introductory section that discusses web-search in general, the importance of image search, and the ranking of search results. Section \ref{sec:fairness} summarizes the bias and fairness aspects of search, and challenges in assessing the fairness of image search outputs. We then discuss different categories of biases in image search in Section \ref{sec:type_biases} and cover occupational stereotyping as an important fairness issue in Section \ref{sec:occupation_stereotyping}. Section \ref{sec:existing_approaches} is devoted to prior arts on mitigation of biases in web search. In section \ref{sec:autoassessment}, we present implementation details of models and frameworks for the assessment of gender stereotypes in image search. Based on this and existing work on text-search re-ranking, we devise a fairness-aware re-ranking algorithm, which is discussed in Section \ref{sec:reranking}. Section \ref{sec:conclusion} concludes the paper with pointers to future work. 
\section{Introduction}
\label{sec:intro}
Search engines play a significant and decisive role in accessing the digital information ecosystem. Every minute, an estimated 3.8 million queries are processed by the Google search engine and this number continues to increase exponentially \cite{grind2019google}. Search engines are arguably the most powerful lines of computer programs in the global economy that controls how much of the world accesses information on the internet. A 2017 international survey found that 86 \% of people use search engines daily \cite{dutton2017search}. Other findings from a study \cite{dutton2013cultures} include that search engines are one of the first places people seek to get information. Moreover, search engines are the second most common use of the internet after email \cite{dutton2017search}. A vast majority of internet-using adults in countries like the U.S.A also rely on search engines to \emph{find and fact-check} information \cite{dutton2013cultures}. A study \cite{mitchell2017americans} shows that search engines are the second-best most likely news gateway that inspires follow-up actions like further searching, online sharing, and talking about the information with others. Hence, search engines have not only attained remarkable growth and usage over a relatively shorter period, but they are also currently proving to be the most trusted source of information \cite{robertson2018auditing}.

% In a recent study by Ka{\v{n}}ukov{\'a} et al. \cite{kavnukova2019searching}, 100\% of participants reported that “doing your own research” started with a Google Search, with one participant believing that Google “works as a fact-checker”. 

The motives behind using a search engine are different for every user.  Users create search terms differently based on their \textit{intentions} and likewise expect different results:  articles, videos, or even an entire site. Even though queries may not always have unique purposes and outcomes, according to Broder et al., \cite{broder2002taxonomy}, there are three basic types of search \textit{i.e.,} (a) \textit{informational search queries} (where the user looks for certain information), (b) \textit{navigational search queries} (when the user wants to visit a specific site or finds a certain vendor), and (c) \textit{transactional search queries} (when the user wishes to execute a transaction– for example buying something). Over the years, the proliferation of internet and web search usage has increased the volume of informational queries amongst other forms by orders of magnitude; this has also given rise to multi-modal search platforms for serving queries for images, speech (audio), web pages, knowledge-cards \textit{etc.}

\subsection{Importance of Image Search}
Images convey much more information as compared to words. They have a powerful impact on what we recognize and what we remember in the future. At any point in time, they speak louder than words. The web tool Mozcast shows that more than 19\% of google searches return images, which means images, rather than texts, are becoming the language of the internet. The growth in visual image search has given rise to a lot of research work in the field of \textit{image information retrieval} (IIR). 

% However, the large number of attributes in images has given rise to various challenges. In this regard, search engines like Google combine data from images, data from the main website, and through deep learning combine similarities within images to provide the best search results for their users.

% \textcolor{red}{TODO: Will add Some excepts from https://sci-hub.se/https://dx.doi.org/10.1109/ICEEOT.2016.7755004}

\subsection{Importance of Search Result Ranking}
\label{sec:intro_importance}
Internet search rankings are known to have a denoting impact on the users' perceptions and decisions, mainly because most of the search users trust and choose higher-ranked results more than lower-ranked results \cite{epstein2015search} and often do not look below the third result \cite{fidel2012human}. Surprisingly, even if the high-ranked results are valued the most, there are no standard qualifiers to identify the top results as the most relevant information for a search keyword \cite{mai2016looking}. Due to the search engine's proprietary nature, users are unaware of the working of its algorithms. The majority of search engine users consider search engine results to be unbiased and accurate \cite{zickuhr2012libraries}. Highly ranked results not only shape a user's opinion and impact his beliefs and unconscious bias; they can also affect his search interactions and experiences. In the same context, it is difficult to believe that these results can often be unfair (biased) in terms of gender, language, demography, socio-cultural aspects, and stereotypes. The bias term is very much attached to the search engines for what they index, what they present overall, and what they present to a particular user. This is a deep concern as people are more vulnerable to bias especially when they are unaware of the biases. Hence, the romanticized view of the search engine that the search engine bypasses the structural biases and skewed data does not match the reality at all. The influence carried by the design decisions of the search engines is broad; it does not only impact the perception of individual information seekers, of society at large it influences our cultures and politics by steering peoples' perspectives towards stereotypical skewed results \cite{robertson2018auditing}.

% Some studies suggest that people will rely on their prior knowledge and select documents that confirm their biases and stereotypes about a particular keyword that they search for \cite{chapman2002incorporating, evans1996rationality}. Cognitive biases and heuristics affect the way people perceive and process new information about a topic – particularly when the learner has to process conflicting or non-intuitive information \cite{tversky1974judgment}. 

\section{Fairness in Web Search}
\label{sec:fairness}
Fairness in web search is the absence of any prejudice or inclination toward an individual or a group based on their inherent or acquired characteristics. In most of the current search engines, there is clear evidence of the absence of fairness that is spread across all the different dimensions of search, i.e., text, image, audio, and speech. If we talk about an image search, there are biased associations between the attributes of an image with representations of social concepts. For instance, the state-of-the-art unsupervised models trained on popular image datasets like ImageNet automatically learn bias from the way that a group of people is stereotypically portrayed on the web \cite{deng2009imagenet}. With the proliferation of artificial intelligence, the Internet of things, and web search and intelligence capabilities in day-to-day life, reducing (if not eliminating) bias is of paramount importance.

\subsection{Examples of Unfairness in Image Search}
A google search result in 2016 for the keyword "three white teenagers" spat out happy and shiny men and women laughing and holding sports equipment. However, the search results for "three black teenagers" offered an array of mug shots. Google acknowledged this bias and responded that the search algorithms mirror the availability and frequency of the online content.``This means that sometimes unpleasant portrayals of sensitive subject matter online can affect what image search results appear for a given query,'' the company said in a statement to the Huffington Post UK. \cite{guarino2016google}

The presentation of black women being sassy and angry presents a disturbing portrait of black womanhood in modern society. In \textit{Algorithms of Oppression}, Safiya Umoja Noble challenges the claim of Google having equity in all forms of ideas, identities, and activities. She argues that the search algorithms that are privileged towards whiteness and discriminate against people of color, essentially women of color, are due to two main factors i.e., (a) Monopoly of a relatively small number of internet search engines and (b) Private interests in promoting certain aspects of the images, which are typically made available when a cursor hovers on the result.

\subsection{Challenges in Evaluating Image Search Fairness vis-\'a-vis General Web Search}
%\subsection{Additional Challenges to evaluation metrics in Image Search as compared to general web search}
Image search results are typically displayed in a grid-like structure unlike that of web search results which are arranged as a sequential list. Users can view/scroll results not only in the vertical direction but in the horizontal direction too. These differences in user behavior patterns lead to challenges in evaluating the search results from a user experience standpoint. There are three key differences in Search Engine Result Pages (SERP) of web search and image search. i.e., (1) An image search engine typically places results on a grid-based panel rather than in a one-dimensional ranked list. As a result, users can view results not only vertically but also horizontally. (2) Users can view results by scrolling down without a need to click on the “next page” button because the image search engine does not have an explicit pagination feature. (3) Instead of a snippet, i.e., a query-dependent abstract of the landing page, an image snapshot is shown together with metadata. \cite{xie2019grid}. These subtle changes in user experience in image search means that users have instant access to more images (\textit{i.e.,} more number of search result outcomes), and because images provide instant access to a large amount of information (as opposed to text/web-pages), tackling biases/unfairness in image search results becomes even more important.
The following section enlists some of the biases that are typically observed in an image search. Note that, we use the terms bias and fairness interchangeably, considering that unfairness could result from certain kinds of biases in search algorithms and procedures.

\section{Type of Image Search Biases}
\label{sec:type_biases}
In this section, we report certain kinds of biases that are typically seen in image search results. 
\begin{enumerate}
    \item \textbf{Position Bias}: One of the key sources of bias in web search results is due to the probability of click which is influenced by a document's position in the SERP(Search Engine Results Page) \cite{craswell2008experimental}.
   % \item \textbf{Societal Bias:} 
   % \item \textbf{Topical Diversity Bias}
    \item \textbf{Confirmation Bias:} For most people, the psychological tendency of interpreting information in web search results has a common ground. They focus on information that confirms their preconceptions. The SERP commonly presents messages with diverse perspectives and expertise, all focused on a single topic or search term. Search results are perhaps unique in the extent to which they can highlight differing views on a topic. Individual convictions lead to one-sided information processing. If these convictions are not justified by evidence, people run into the risk of being misinformed \cite{schweiger2014confirmation}.
    \item \textbf{Domain Bias:} Ieong et al. [24] investigated domain bias, a phenomenon in Web search that users’ tendency to prefer a search result just because it is from a reputable domain, and found that domains can flip a users preference about 25\% of the time under a blind domain test.
    \item \textbf{Selection Bias:} This bias occurs when a dataset is imbalanced for different regions or groups; it over-represents one group and under-represents the other. When ML algorithms are trained through web-scraping, the search results mostly revolve around the data that are present in a vast amount on the web. So, the selection does not reflect the random sample and is not representative of the actual population here. This particular bias, which could be referred to as the (re)search bubble effect, is introduced because of the inherent, personalized nature of internet search engines that tailors results according to derived user preferences based on non-reproducible criteria. In other words, internet search engines adjust their user’s beliefs and attitudes, leading to the creation of a personalized (re)search bubble, including entries that have not been subjected to a rigorous peer-review process. The Internet search engine algorithms are in a state of constant flux, producing differing results at any given moment, even if the query remains identical \cite{curkovic2018bubble}.
    \item \textbf{Historical Bias:} This type of bias comes from socio-economic issues in the world and passes on gradually right from the data generation process. For example, while searching for images of nurses, only a few male nurses show up in the search result. This type of bias occurs due to the presence of already existing stereotypes based on historical data. Suppose for a profession, most of the persons were male earlier. But now people of both genders work in that profession. If the data at a given time frame characterizes the creator's preconceived notions, it may result in historical bias as time progresses \cite{lim-etal-2020-annotating}.
    \item \textbf{Human Reporting Bias:} The frequency of a particular type of content may not be a reflection of the real-world frequencies of that content/event. What people share on the web may not be a reflection of real-world frequencies. For example, wedding images from various cultures may not be uniformly uploaded to the web and indexed \cite{kulshrestha2017quantifying}.
    \item \textbf{Racial Bias:} This form of bias occurs when data skews in favor of particular demographics. For instance, indexing a greater number of images of certain demography (\textit{e.g.,} younger population or races) is prevalent in certain countries just because this demographic population accesses and uses the web applications more \cite{kulshrestha2017quantifying}.
    \item \textbf{Association bias:} This bias occurs when the data for a machine learning model multiplies the cultural bias. For instance, datasets created in an automatic/semi-automatic manner may have a collection of jobs in which all men are doctors, and all women are nurses. This does not necessarily mean that women cannot be doctors, and men cannot be nurses \cite{lim-etal-2020-annotating}.
\end{enumerate}

Various forms of biases undoubtedly can be a potential menace for the active internet population. Some kinds of bias can even have a more adverse impact and should be mitigated/addressed both from the dimensions of a system (algorithmic) and user (behavioral). In the following section, we delve into one such form of bias \textit{i.e.,} occupational stereotypes, a specific yet important fairness issue in image search results.
\section{Occupational Stereotypes in Image Search Results}
\label{sec:occupation_stereotyping}
Stereotyping is the generalization of a group of people. At times even if it is statistically almost accurate, it is not universally valid. In this context, one of the most prevalent and persistent biases in the United States is portraying and perpetuating inequality in the representation of women on various online information sources \cite{zhao2018gender}. 

A recent study from the University of Washington\cite{langston2015sa} assessed the gender representations in online image search results for 45 different occupations. The study founds that in a few jobs like CEO, women are significantly underrepresented in Google search results. This study also claims that across all other professions too, women are slightly underrepresented on average. Other search results data published \cite{silberg2019notes} also show similar trends - for example, the percentage of women in the top 100 Google image search results for CEO is 11 percent in contrast to the actual percentage of women CEOs in the US which is 27 percent. These biases are highly insidious as they are neither transparent to the user nor the search engine designers.

This form of bias is mainly attributed to two main factors \textit{i.e.,} (a) Slight exaggeration of gender ratios and (b) Systematic over or under-representation of genders \cite{kay2015unequal}. Male-dominated professions have even more men in the search results than it is supposed to have(real-world distributions). This effect is even prevalent when people rate the quality of search results or select the best image that represents an occupation. They unknowingly prefer the image with a gender that matches the stereotype of that particular occupation. 

\begin{figure}[t]
\captionsetup[subfigure]{labelformat=empty}
    %\centering
\begin{subfigure}{0.20\textwidth}
\subfloat[0, Female]{%
  \includegraphics[width=0.9\linewidth, height = 2 cm]{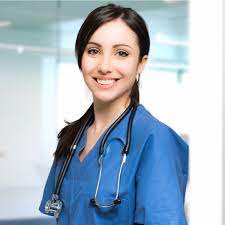}}
  %\caption{Rank 0}
  \label{Rank0_Nurse}
\end{subfigure}\hfil
\begin{subfigure}{0.20\textwidth}
\subfloat[1, Both]{%
  \includegraphics[width=0.9\linewidth, height = 2 cm]{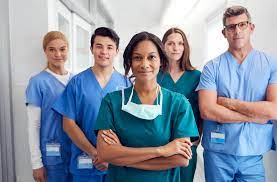}}
  %\caption{Rank1_Nurse}
  \label{Rank1_Nurse}
\end{subfigure}\hfil
\begin{subfigure}{0.20\textwidth}
\subfloat[2, Female]{%
  \includegraphics[width=0.9\linewidth, height = 2 cm]{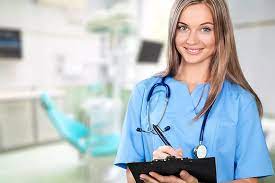}}
  %\caption{Rank 4}
  \label{Rank2_Nurse}
\end{subfigure}\hfil
\begin{subfigure}{0.20\textwidth}
\subfloat[3, Both]{%
  \includegraphics[width=0.9\linewidth, height = 2 cm]{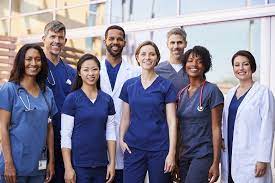}}
  %\caption{Rank 4}
  \label{Rank3_Nurse}
\end{subfigure}\hfil
\begin{subfigure}{0.20\textwidth}
\subfloat[4, Female]{%
  \includegraphics[width=0.9\linewidth, height = 2 cm]{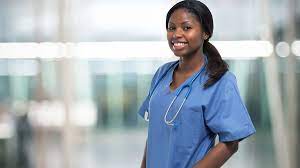}}
  %\caption{Rank 4}
\end{subfigure}\hfil
\medskip
\begin{subfigure}{0.20\textwidth}
\subfloat[5, Both]{%
  \includegraphics[width=0.9\linewidth, height =2 cm]{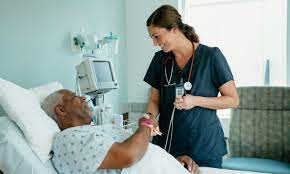}}
  %\caption{Rank 5}
\end{subfigure}\hfil
\begin{subfigure}{0.20\textwidth}
\subfloat[6, Female]{%
  \includegraphics[width=0.9\linewidth, height = 2 cm]{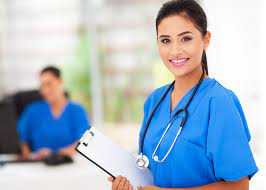}}
  %\caption{Rank 6}
\end{subfigure}\hfil
\begin{subfigure}{0.20\textwidth}
\subfloat[7, Female]{%
  \includegraphics[width=0.9\linewidth, height = 2 cm]{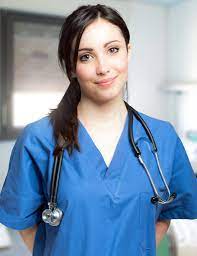}}
  %\caption{Rank 7}
\end{subfigure}\hfil
\begin{subfigure}{0.20\textwidth}
\subfloat[8, Both]{%
  \includegraphics[width=0.9\linewidth, height = 2 cm]{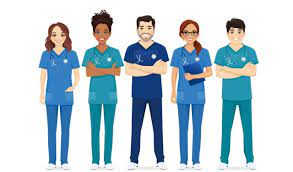}}
  %\caption{Rank 8}
\end{subfigure}\hfil
\begin{subfigure}{0.20\textwidth}
\subfloat[9, Female]{%
  \includegraphics[width=0.9\linewidth, height = 2 cm]{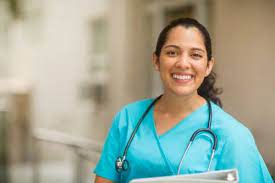}}
  %\caption{Rank 9}
\end{subfigure}\hfil
\caption{Distribution of genders across top 10 Google Search results for the query term "Nurse", as of June, 2021}
\label{fig:gen_dist1}
\end{figure}

\begin{figure}[t]
\captionsetup[subfigure]{labelformat=empty}
    %\centering
\begin{subfigure}{0.20\textwidth}
\subfloat[0, Male]{%
  \includegraphics[width=0.9\linewidth, height = 2 cm]{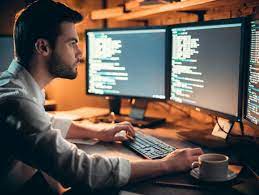}}
  %\caption{Rank 0}
  \label{Rank0_Engineer}
\end{subfigure}\hfil
\begin{subfigure}{0.20\textwidth}
\subfloat[1, Uncertain]{%
  \includegraphics[width=0.9\linewidth, height = 2 cm]{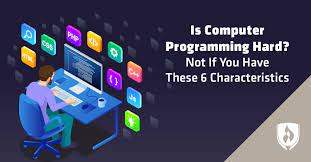}}
  %\caption{Rank1_Engineer}
  \label{Rank1_Engineer}
\end{subfigure}\hfil
\begin{subfigure}{0.20\textwidth}
\subfloat[2, Female]{%
  \includegraphics[width=0.9\linewidth, height = 2 cm]{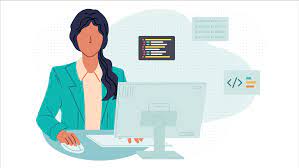}}
  %\caption{Rank 4}
  \label{Rank2_Engineer}
\end{subfigure}\hfil
\begin{subfigure}{0.20\textwidth}
\subfloat[3, Uncertain]{%
  \includegraphics[width=0.9\linewidth, height = 2 cm]{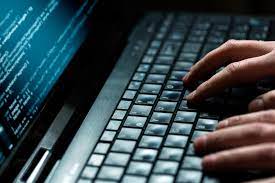}}
  %\caption{Rank 4}
  \label{Rank3_Engineer}
\end{subfigure}\hfil
\begin{subfigure}{0.20\textwidth}
\subfloat[4, Male]{%
  \includegraphics[width=0.9\linewidth, height = 2 cm]{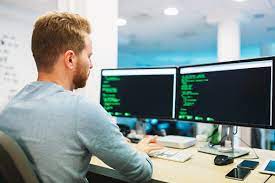}}
  %\caption{Rank 4}
\end{subfigure}\hfil
\medskip
\begin{subfigure}{0.20\textwidth}
\subfloat[5, Female]{%
  \includegraphics[width=0.9\linewidth, height = 2 cm]{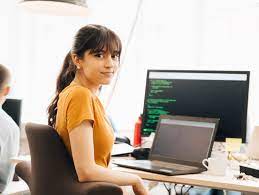}}
  %\caption{Rank 5}
\end{subfigure}\hfil
\begin{subfigure}{0.20\textwidth}
\subfloat[6, Male]{%
  \includegraphics[width=0.9\linewidth, height = 2 cm]{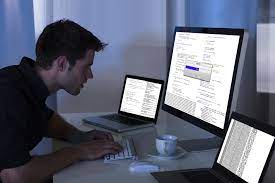}}
  %\caption{Rank 6}
\end{subfigure}\hfil
\begin{subfigure}{0.20\textwidth}
\subfloat[7, Male]{%
  \includegraphics[width=0.9\linewidth, height = 2 cm]{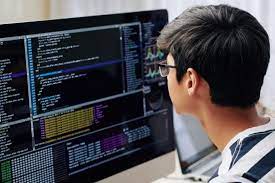}}
  %\caption{Rank 7}
\end{subfigure}\hfil
\begin{subfigure}{0.20\textwidth}
\subfloat[8, Both]{%
  \includegraphics[width=0.9\linewidth, height = 2 cm]{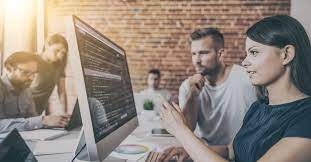}}
  %\caption{Rank 8}
\end{subfigure}\hfil
\begin{subfigure}{0.20\textwidth}
\subfloat[9, Male]{%
  \includegraphics[width=0.9\linewidth, height = 2 cm]{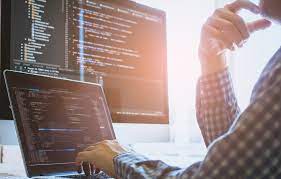}}
  %\caption{Rank 9}
\end{subfigure}\hfil
\caption{Distribution of genders across top 10 Google Search results for the query term "Computer Programmer" as of June 2021}
\label{fig:gen_dist2}
\end{figure}

While ranking images in search results based on quality, people do not systematically prefer either gender. However, here stereotyping dominates in the decision-making process. They prefer images with the gender that matches the stereotype for that occupation. Additionally, the skewed search results also exhibit biases in how genders are depicted overall. The results that match with the gender stereotype of a profession tend to be portrayed as more professional-looking and less inappropriate-looking \cite{kay2015unequal}. Figures \ref{fig:gen_dist1} and \ref{fig:gen_dist2} provide insights into such issues and disparities for frequent occupational image search queries.


For male-dominated professions, both of the aforementioned effects i.e., a slight exaggeration of gender ratios and systematic over or under-representation of genders amplify each other. In female-dominated professions, these two effects cancel each other \cite{kay2015unequal}. The study also revealed that there may be a slight under-representation of women and there may be a slight exaggeration of gender stereotypes, but it’s not completely different or divorced from reality. As research and strategy development nonprofit \textit{Catalyst} reports, women currently hold only 30 (\textit{i.e.}, 6\%) CEO positions at S\&P500 companies \cite{hopkins2021buried}.

In 2015, a journalist writing about this study found that when searching for CEOs, the first picture of a woman to appear on the second page of image results was a "Barbie doll" \cite{hopkins2021buried}. Furthermore, another study revealed that Google’s online advertising system features advertisements for high-income jobs for male Internet users much more often than female users \cite{schroeder2015critical}.

Some other examples of occupational stereotypes are as follows. The image search for the keyword "US authors" results in only twenty-five percent of women in the search results in contrast to the actual percentage of 56\%. Similarly, the search results for the keyword "telemarketers" depict 64\% female pictures. However, that occupation is evenly split between men and women in real \cite{langston2015sa}.

In one research study from the University of Washington, the participants were to rank the images based on the professionalism depicted in top image results. This study found that the majority gender for a profession tended to be ranked as more competent, professional, and trustworthy \cite{langston2015sa}. The images of persons whose gender does not match with the occupational stereotype are more likely to be rated as provocative or inappropriate, \textit{e.g.}, Construction Workers. "Getty Images last year created a new online image catalog of women in the workplace – one that countered visual stereotypes on the internet of moms as frazzled caregivers rather than powerful CEOs" \cite{schroeder2015critical}.

To name this as a problem, we need to understand whether this gender stereotyping affects or shifts the users’ perceptions regarding the dominance of gender in that particular profession. The results from a study by the University of Washington researchers \cite{kay2015unequal} hint that the exposure to the skewed image search results shifted the perceptions of users by 7\% at least for short-term changes in perceptions. However, these short-term biases over time can have a lasting effect, starting from personal perceptions to the high-valued decision-making process like hiring.

The skewed representation and gender stereotypes in image search results for occupations also contribute to the type of images selected by users. An image that matches the stereotype for an occupation is more likely to be selected as an exemplar result \cite{kay2015unequal}.

From the aforementioned points, we are certain that this occupational stereotype has an adverse effect on altering users' belief systems about different occupations and their related attributes. So, there is a need to mitigate this type of bias in image search results. Some of the existing approaches to alleviate this bias and our own implementation for de-biasing are described in the following sections.

% Entrepreneurs are crucial to the vitality of an economy. The information that people access affects their perceptions of the world and significantly affects the decisions regarding the choices and opportunities they make in day-to-day life. In this setting, the measured effect of gender stereotypes in image search results raises an interesting question regarding whether the search algorithms need to be modified to counter this skewed representation of gender occupational stereotypes.

\section{Existing Approaches to Mitigate Bias in Image Search}
\label{sec:existing_approaches}
Due to the growing dependence on search engines, automatic curation of biased content has become a mandate. Mitigating biases in search results would promote effective navigation of the web and will improve the decision-making of users \cite{fogg2002persuasive}. One possible way is to collect a large number of ranked search results and re-rank them in a \textit{post-hoc} manner so that the top results shown to the users become fairer. Another possible way is to make changes in the search and ranking algorithms so as to address biases. Post-hoc approaches are model agnostic and hence, more preferable in information retrieval, especially considering the black-box nature of search systems.

\subsection{Mitigation through Re-ranking of Search Results}
On an unprecedented scale and in many unexpected ways, surprisingly search ranking makes our psychological heuristics and vulnerabilities susceptible \cite{bond201261}. Algorithms trained on biased data reflect the underlying bias. This has led to emerging of datasets designed to evaluate the fairness of algorithms and there have been benchmarks to quantify discrimination imposed by the search algorithms \cite{hardt2016equality,kilbertus2017avoiding}. In this regard, the goal of re-ranking search results is to bring more fairness and diversity to the search results without the cost of relevance. However, due to severely imbalanced training datasets, the methods to integrate de-biasing capabilities into these search algorithms still remain largely unsolved. Concerns regarding the power and influence of ranking algorithms are exacerbated by the lack of transparency of the search engine algorithms \cite{pasquale2015black}. Due to the proprietary nature of the system and the requirement of high-level technical sophistication to understand the logic makes parameters and processes used by these ranking algorithms opaque\cite{pasquale2015black, gillespie2014relevance}. To overcome these challenges, researchers have developed techniques inspired by social sciences to audit the algorithms to check for potential biases \cite{sandvig2014auditing}. To quantify bias and compute fairness-aware re-ranking results for a search task, the algorithms would seek to achieve the desired distribution of top-ranked results with respect to one or more protected attributes like gender and age\cite{epstein2017suppressing}. This type of framework can be tailored to achieve such as equality of opportunity and demographic parity depending on the choice of the desired distribution \cite{geyik2019fairness}.

%\subsection{There is a need of improving the performance of keyword-based image search engines by re-ranking their original results. Why?}
\subsection{Need of re-ranking keyword-based image search results}
%\subsubsection{Current limitations of keyword-based image search}
According to a study \cite{jain2011learning}, there are three limitations for keyword-based image search \textit{i.e.,}
\begin{enumerate}[label=(\alph*)]
    \item There is no straightforward and fully automated way of going from text queries to visual features. In a search process, visual features are mainly used for secondary tasks like finding similar images. Since the search keyword/query is fed as a text, rather than an image to a search engine, the search engines are forced to rely on static and textual features extracted from the image’s parent web page and surrounding texts which might not describe its salient visual information.
    \item Image rankers are trained on query-image pairs labeled with relevance judgments determined by human experts. Such labels are well known to be noisy due to various factors including ambiguous queries, unknown user intent, and subjectivity in human judgments. This leads to learning a sub-optimal ranker.
    \item A static ranker is typically built to handle disparate user queries. Therefore, the static ranker is unable to adapt its parameters to suit the query at hand and it might lead to sub-optimal results. In this regard, Jain et al.\cite{jain2011learning} demonstrated that these problems can be mitigated by employing a re-ranking algorithm that leverages \textit{aggregate user click-through data}.
    
    % For this, a  static ranker is learned for all query classes within a vertical and the learned ranker does not need to adapt its parameters to cope with the very diverse set of user queries.
\end{enumerate}

There are different types of methods for re-ranking keyword-based image search results and those are described in the following subsections.

\subsection{Reranking by Modeling Users' Click Data}
One way to re-rank the search engine results is through the user click data. For a given query if we can identify images that have been clicked earlier in response to that query, a Gaussian Process (GP) regressor is trained on these images to predict their normalized click counts. Then, this regressor can be used to predict the normalized click counts for the top-ranked 1000 images. The final re-ranking would be done based on a linear combination of the predicted click counts as well as the original ranking scores \cite{epstein2015search}.
%\subsubsection{How does the above-mentioned model tackle the aforementioned problems and still coped with the limitations mentioned above?}
This way of modeling tackles re-ranking nicely while still coping with the limitations described earlier.
\begin{itemize}
    \item The GP regressor is trained on not just textual features but also visual features extracted from the set of previously clicked images. Consequently, images that are visually similar to the clicked images in terms of measured shape, color, and texture properties are automatically ranked high.
    \item Expert labels might be erroneous, inconsistent with different experts assigning different levels of relevance to the same query-image pair. Such factors bias the training set that results in the learned ranker being sub-optimal. The click-based re-ranker provides an alternative by tackling this problem directly. The hypothesis is that, for a given query, most of the previously clicked images are highly relevant and hence should be leveraged to mitigate the inaccuracies of the baseline ranker.
    \item As the GP regressor is trained afresh on each incoming query, it is free to tailor its parameters to suit a given query at hand. For example, images named \textit{TajMahal.jpg} are extremely likely to be of the Taj Mahal. For landmark queries, the query-image file-name match feature is important. But this feature may be uninformative for city queries. For example - Images of Delhi’s tourist attractions are sometimes named \textit{delhi.jpg}. Also, people's photographs during their trip to Delhi may be named \textit{delhi.jpg}. In this case, a single static ranker would be inadequate. However, the GP regressor aims to learn this directly from the click training data and weights this feature differently in these two situations. 
\end{itemize}
% \subsubsection{Why should the key assumption work?}
Here the key assumption is that "For a given query the clicked images are highly relevant". It would work for an image search due to an obvious reason. For a normal textual web search, only a two-line snippet for each document is displayed in the search results. So, the clicked documents might not be relevant to the keyword. The relevance can only be determined if the user goes or does not go through the document. However, in the case of an image search, most search results are thumbnails allowing users to see the entire image before clicking on it. Therefore, the user predominantly tends to click on the relevant images and most likely discards distracting images\cite{jain2011learning}.

\subsection{De-biased Reinforcement Learning Click model (DRLC) for re-ranking search results}
The users' clicks on web search results are one of the key signals for evaluating and improving web search quality, hence is widely used in state-of-the-art \textit{Learning-To-Rank}(LTR) algorithms. However, this has a drawback from the perspective of fairness of the ranked results. These algorithms can't justify the scenario when a search result is not necessarily clicked as it is not relevant. Rather it is not chosen because of the lower rank assigned to it on the SERP. If this kind of bias in the users' click log data is incorporated into any LTR ranking model, the underlying bias would be propagated to the model. In this regard, a reinforcement learning model for re-ranking seems to be very effective and can avoid the proliferation of position bias in the search results.

\subsubsection{Importance of Reinforcement Learning in Information Retrieval Process}
In today's world of advanced modern information retrieval interface which typically involves multiple pages of search results, the users are likely to access more than one page. In a common retrieval scenario, the search results are split into multiple pages that the user traverses across by clicking the "next page" button. The user generally believes in the ranking of the Search Engine Results Page(SERP) and examines the page by browsing the rank list from top to bottom; he clicks on the relevant documents and returns to the SERP in the same session. In the case of a good multi-page search system, it begins with a static method and continues to adopt a model based on feedback from the user. Here comes the importance of reinforcement learning. This type of relevance feedback method\cite{joachims2007evaluating} has been proven to be very effective for improving retrieval accuracy over interactive information retrieval tasks.

According to the Rocchio algorithm\cite{rocchio1971smart}, the search engine gets feedback from the user; adds weights to the terms from known relevant documents, and minus weights of the terms from the known irrelevant documents. So, most of the feedback methods balance the initial query and the feedback information based on a fixed value.

Learning to rank methods have been widely used for information retrieval, in which all the documents are represented by feature vectors to reflect the relevance of the documents to the query \cite{liu2011learning}. The learning-to-rank method aims to learn a score function for the candidate documents by minimizing a carefully designed loss function. The work from Zeng et al. \cite{zeng2018multi} considers the multi-page search scenario and applies \textit{relevance feedback} techniques to state-of-art learning to rank models. The multi-page search process is an interactive process between the user and the search engine. At each time step, the search engine selects M documents to construct a rank list. The user browses this rank list from top to down, clicks the relevant documents, skips the irrelevant documents, and then clicks the "next page" button for more results. In the paper \cite{zeng2018multi}, multi-page search processes are mathematically formulated as a Markov Decision Process. The search engine is treated as the agent, which selects documents from the remaining candidate document set to deliver to the user for satisfying the user’s information need. The state of the environment consists of the query, remaining documents, rank position, and user’s click information. The soft-max policy is applied to balance the exploration and exploitation during training and design the reward based on the IR measure metric. A classical policy gradient policy method based on the REINFORCE algorithm is applied to optimize the search policy.

\begin{figure}[t]
    \centering
    \includegraphics[width=12cm]{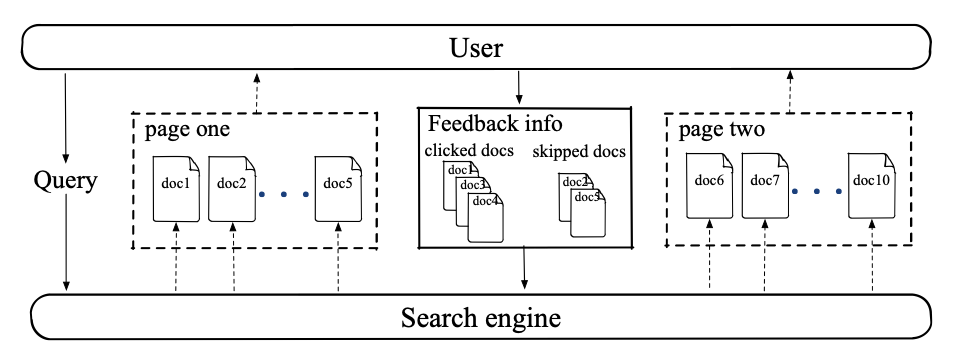}
    \caption{Multi Page Search Process by Zeng et al. \cite{zeng2018multi}}
    \label{fig:multi_stage}
\end{figure}
Zeng et.al. \cite{zeng2018multi}, proposed a technical schema to use user feedback from the top-ranked documents to generate re-ranking for the remaining documents. Compared with existing methods, their method enjoys the following advantages: (i) it formulates the multi-page process as a Markov Decision Process and applies policy gradient to train the search policy which could optimize the search measure metric directly, (ii) it applies the Recurrent Neural Network to process the feedback and improve the traditional learning to rank model with the feedback information based on Rocchio. The authors used traditional learning to rank the method ListNet, RankNet, and RankBoost as the initial model to construct the experiments on the OHSUMED dataset and simulate the interaction between the search engine and user based on Dependent Click Model (DCM). Experimental result shows that their model can prove the ranking accuracy for traditional learning to a rank method and has better generalization ability.

Zhou et al. \cite{zhou2021biased} proposed a De-biased Reinforcement Learning Click model (DRLC) that ignores previously made assumptions about the users' examination behavior and resulting latent bias. To implement this model, CNNs are used as the \textit{value network} for reinforcement learning, trained to log a policy to reduce bias in click logs. The experiments demonstrated the effectiveness of the DRLC model in learning to reduce bias in click logs, leading to improved modeling performance and showing the potential for using DRLC for improving Web search quality.

It is worth mentioning that probabilistic Graphical Model Frameworks (GMFs) have been traditionally used for search result ranking and there are basically two broad ways these models operate in: 
\begin{enumerate}[label=(\alph*)]
    \item \textbf{Direction 1:} This considers the search process as a sequence of events. Predicting clicks is based on some probability models and assumptions. While they are flexible and interpreted, this is limited by a weak learning model with fewer features.
    \item \textbf{Direction 2:} This model considers the searching process as represented by some vectors. While this model allows the users to consider a variety of features easily and feed them to a stronger learning model such as neural nets \cite{chakraborty2000bishop}, this model can not consider the bias issue in an interpretable way. 
\end{enumerate}

As DRLC is also kind of a PGM-based method, it can be organized in a flexible way for different ranking scenarios and generate an interpretive model to reduce a variety of biases. DRLC is constructed by a more dynamic system, which is the reinforcement learning \cite{sutton2018reinforcement, zhou2020rlirank}  This allows DRLC to take advantage of stronger learning models (NNs). DRLC model can thus overcome issues faced by traditional GMFs.
\subsection{Best Practices and Policies to Mitigate Search Engine's, Algorithmic Bias}
According to Lee et al. [30], understanding various causes of biases is the first step for adopting effective algorithmic hygiene. But it is a challenging task to assess the search results for bias. Even when flaws in the training data are corrected, the results may still be problematic because context matters during the bias detection phase. When detecting bias, computer programmers generally examine the set of outputs that the algorithm produces to check for anomalous results. However, the downside of these approaches is that not all unequal outcomes are unfair; even error rates are not a simple litmus test for biased algorithms. In this regard, below are some of the evaluation protocols and metric formulations.
\begin{itemize}
    \item Conducting quantitative experimental studies on bias and unfairness.
    \item Defining objective metrics that consider fairness and/or bias.
    \item Formulating bias-aware protocols to evaluate existing algorithms.
    \item Evaluating existing strategies in unexplored domains.
\end{itemize}

In the following sections, we discuss our implementations and insights on assessing occupational stereotypes in image search results and also debasing the search results through re-ranking. As gender stereotyping seems to be a prevalent issue in occupational search results, we explore various techniques for detecting gender distributions of given sets of images (which correspond to a set of retrieved images from a search engine). This is presented in Section \ref{sec:autoassessment}. After this, in Section \ref{sec:fairness}, we present a framework for reranking image search results to make the results fairer while preserving the relevance of the images retrieved with respect to the input query.
\begin{table}[t]
\centering
\footnotesize
\begin{tabular}{ l l l l l l l }
\hline
\hline
\textbf{Name} & \textbf{Source} & \textbf{Framework} & \textbf{Input} & \textbf{Output} & \textbf{Size} \\
\hline
SSR-Net \cite{yang2018ssr} & \url{t.ly/MFwu} & Keras/TensorFlow & (64,64,3) & Real & 0.32Mb \\
ConvNet \cite{LH:CVPRw15:age} & \url{t.ly/FPFT} & Caffe & (256,256,3) & Binary & 43.5Mb \\
Inception-V3 (Carnie)& \url{t.ly/SxXJ} & TensorFlow & (256,256,3) & Binary & 166Mb \\
ConvNet (Jiang) & \url{t.ly/IMSY} & TensorFlow & (160,160,3) & Real & 246Mb \\
ConvNet (Chengwei) & \url{t.ly/G9ff} & Keras/TensorFlow & (64,64,3) & Real & 186Mb \\
ConvNet (Serengil) & \url{t.ly/6WH9} & Keras/TensorFlow & (224,224,3) & Real & 553Mb \\
\hline 
\end{tabular}
\caption{Existing approaches to gender detection from facial data}
\label{tab:existing}
\end{table}
\begin{figure}[t]
     \centering
     \includegraphics[width=12cm]{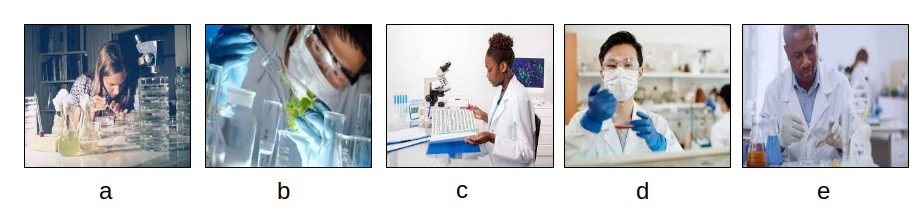}
     \caption{Top 5 search results for the keyword ``biologist'' (circa: June 2021)}
     \label{fig:searchresults}
\end{figure}
\section{\textit{Implementation:} Automatic Assessment Occupational Stereotypes}
\label{sec:autoassessment}
In this section, we investigate and re-implement a few existing frameworks for occupational stereotype assessment of image-search results. One key component of the analysis of the fairness of search results for occupations is to get insights into the distribution of gender in the ranked search results. To this end, the following sections aim to discuss our implementations, experimental results, and assessment of challenges involved in computing the distribution of gender in search results.

\subsection{Automatic Detection of Gender of Image Search Results: Open Source Systems}
\label{sec:genderdetection}
Automatic gender detection from images deals with classifying images into gender categories such as \textit{male}, \textit{female}, \textit{both} and \textit{uncertain}. Of late, gender identification has become relevant to an increasing amount of applications related to facial recognition and analysis, especially since the rise of social media. Gender detection typically relies on spatial features related to the human body, specifically the face, which needs identification and alignment of facial elements in the image. The problem becomes harder when the environment becomes unconstrained, such as in the case of web-search retrieved images. This is primarily due to variations in poses, illuminations, occlusions, and interference of other objects with the facial region of interest. Figure \ref{fig:searchresults} show this through top-5 search results obtained from a popular search engine for the keyword ``biologist''. As it can be seen, the result images can be diverse and very different from frontal face images typically used by applications requiring gender identification. 

In the following section, we summarize some of the existing works on this problem and enlist a few \textit{ off-the-shelf} methods that can be tried for fairness assessment related to gender.

\subsubsection{Existing Open Source Systems}
\label{sec:gender_releted}
Following the article by Chernov \cite{Ageandge68:online}, we summarize some of the existing projects on gender identification in Table \ref{tab:existing} that can be tried off-the-shelf on image search results. 

Most of the current gender identification model relies on a pre-processing step (details given in Section \ref{sec:gender_idea} of face recognition and several models exists for that as well. Some of the popular implementations are:
\begin{itemize}
    \item \textbf{OpenCV Harcascade:} OpenCV \footnote{\url{https://opencv.org/}} is a collection of APIs for real-time computer vision. It comes up with a powerful cascade classifier for object detection \cite{viola2001rapid} and provides pre-trained machine learning models for frontal face detection. It implements classical machine learning based techniques for classification.  
    \item \textbf{ConvNet by Lavi and Hassen:} Lavi and Hassen  \cite{LH:CVPRw15:age} have released a set of deep learning models for face detection, and gender and age identification. 
    \item \textbf{MultiTask CNN:} Multitask model for joint modeling of face alignment and detection\cite{zhang2016joint} aims to mutually learn face alignment and identification tasks and thus benefit each individual task. 
    \item \textbf{Facelib:} Facelib \footnote{https://github.com/sajjjadayobi/FaceLib} implements MobileNets\cite{howard2017mobilenets} based face detection module. It also has a self-trained gender predictor based on the UTKFace Dataset \cite{zhifei2017cvpr}.  
    \item \textbf{Detectron:} Detectron \cite{wu2019detectron2} is a popular deep learning based tool for object detection and image segmentation. It implements masked Region Based CNNs for object segmentation. Although the pretrained models from Detectron and Detectron2 may not provide segmentation for faces, they can help extract segments corresponding to persons in the images. This is particularly beneficial when the images do not contain clear faces, but partial and oriented human faces and bodies. 
\end{itemize}
\begin{figure}[t]
    \centering
    \includegraphics[width=14cm]{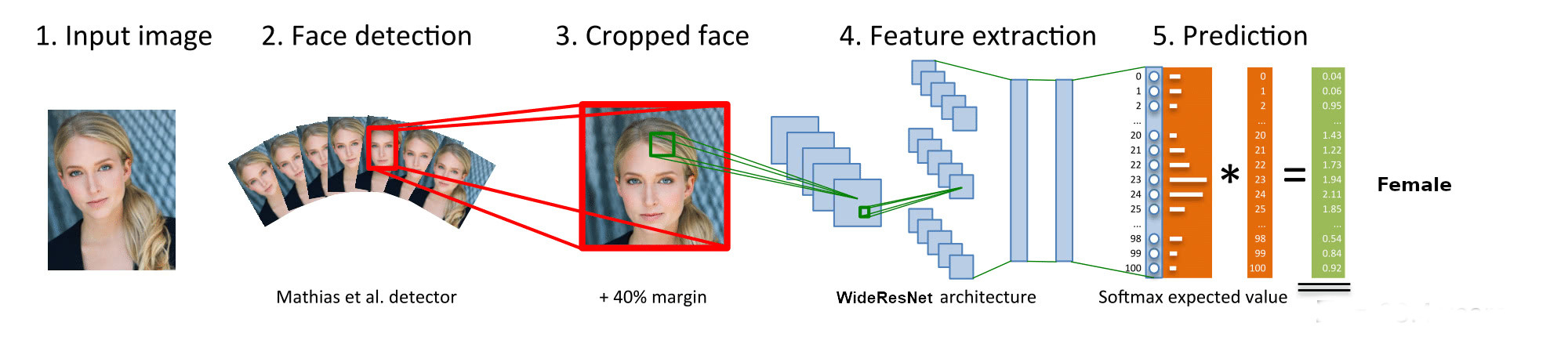}
    \caption{Pipeline architecture\cite{rothe2015dex} commonly followed for gender identification from images}
    \label{fig:pipeline}
\end{figure}
\subsubsection{Central Idea}
\label{sec:gender_idea}
Most of the existing approaches to gender identification from images follow a pipeline approach as shown in Figure \ref{fig:pipeline}. The input image is pre-processed (sometimes gray-scaled) and then passed through a face/body identification model that extracts the region of interest \textit{i.e.}, human face or body. The original image is then cropped to retain the identified portion, which is then given to a gender/age prediction model. The gender prediction model is typically a deep neural network like multi-layered Convolutional Neural Network (CNN), which is responsible for extracting gender-specific semantic representations from the cropped image and pass this information onto a dense layer of a neural classifier.

Since we are interested in conducting inference on search results, we should ideally consider pre-trained models (like the ones described in Section \ref{sec:gender_releted}). It goes without a say that existing models exhibit superior/acceptable performance on the frontal face, as they are often trained on datasets like ImageNet or UTKFace and with frontal face images. Their performance though reduces when the environment becomes unconstrained, such as in the case of web-search retrieved images. In the following section, we discuss some of the challenges related to gender identification web-search retrieved images.
\subsubsection{Challenges in Estimation of Gender using Existing Frameworks}
\label{sec:gender_challenges}
Image search results can result in a truly diverse set of images with varying poses, illuminations, occlusions, artifacts, orientation, and quality. This throws additional challenges for existing pre-trained gender detection frameworks. Some of the challenges we observed in though our challenge dataset (refer Section \ref{sec:gender_dataset}) are given below:
\begin{enumerate}
    \item \textbf{Face / body partially visible and/or misaligned:} Unlike frontal images, images from keyword search may be misaligned / partially informative. For example, Figure \ref{fig:searchresults} (a), (b), (c) do which correspond to top-ranked results do not provide enough information about faces and gender. Specifically,  occupation related searches will retrieve images of persons focusing on their work, which result in a higher number of non-frontal, partial and misaligned images. 
    \item \textbf{Interference of objects:} For occupational key-word search, the retrieved images may have occupation related objects/instruments blocking faces/bodies. For example, biologists are often seen with instruments like microscopes, or wearing masks. 
    \item \textbf{Images with varying degree of quality and resolutions:} Since web search results are nothing but indexed images, they may vary in terms of resolution, frame height and width and aspect ratio. Hence, processing all the retrieved images may not yield ideal results. 
\end{enumerate}
\begin{table}[t]
\centering
%\footnotesize
\begin{tabular}{l l l c c c c }
\hline
\hline
& \textbf{Search Term} & \textbf{Abbr.} & \%M & \%F & \%Both & \%Uncertain\\
\hline
1 & ``biologist'' & BIO & 27 & 43 & 11 & 19 \\
2 & ``chief executive officer'' & CEO & 51 & 11 & 6 & 32  \\
3 & ``cook'' & COOK & 23 & 30 & 13 & 34  \\
4 & ``engineer'' & ENG & 43 & 18 & 21 & 18  \\
5 & ``nurse'' & NUR & 1 & 50 & 43 & 1  \\
6 & ``police officer'' & POL & 70 & 10 & 18 & 2  \\
7 & ``primary school teacher'' & PST & 5 & 11 & 81 & 3 \\
8 & ``computer programmer''  & PRO & 38 & 13 & 13 & 36 \\
9 & ``software developer'' & SD & 26 & 11 & 14 & 49 \\
10 & ``truck driver'' & TD & 44 & 7 & 0 & 49 \\
\hline
\end{tabular}
\caption{Search terms and gender distribution in the collected image search dataset }
\label{tab:datastats}
\end{table}
\subsubsection{Creation of Challenge Image-Search Datasets for Evaluation}
\label{sec:gender_dataset}
From our initial inspection of search results for occupational queries such as ``biologist'', we observed that the images retrieved had significant differences from typical frontal face images used in most of the gender identifiers. We prepare a manually labeled challenge dataset for evaluating gender identification systems, targeted for assessing and de-biasing search systems. The dataset was created by manually assigning gender labels to the top 100 image search results from Google for a specific keyword. We collected search results for 10 occupation keywords, as shown in Table \ref{tab:datastats}. For annotation, we considered 4 labels: (a) Male (b) Female (c) Both (d) Uncertain. For labeling, we leveraged the Amazon Mechanical Turk framework, where workers were asked to assign one of these labels for each image. Appropriate guidelines were given to them to tackle ambiguous cases, especially cases where faces/gender-specific attributes are not clearly visible from the images.

As shown in Table \ref{tab:datastats}, for certain occupations like the nurse, the results are heavily biased towards the female gender, whereas it is completely opposite for a male dominated occupation such as truck driving. It is also interesting to note that, for certain occupations like cook and software developer, the top-ranked images, more often than not, do not provide any gender-specific information.
\begin{table}[t]
\centering
\footnotesize
\begin{adjustwidth}{-0.5in}{-1in}
\begin{tabular}{ l l l l l l l l l l l l}
\hline
\hline
\textbf{Candidate Pipeline} & \textbf{BIO} & \textbf{CEO} & \textbf{COOK} & \textbf{ENG} & \textbf{NUR} & \textbf{POL} & \textbf{PST} & \textbf{PRO} & \textbf{SD} & \textbf{TD} & \textbf{Avg.}\\
\hline
Cascade + Facelib & 0.31 & 0.7 & 0.42 & 0.33 & 0.36 & 0.41 & 0.07 & 0.42 & 0.51 & 0.57 & 0.41 \\
ConvNet + Facelib & 0.39 & 0.81 & 0.55 & 0.38 & 0.19 & 0.53 & 0.19 & 0.48 & 0.54 & 0.64 & 0.47 \\
MTCNN + Facelib & 0.49 & 0.87 & 0.64 & 0.54 & 0.31 & 0.68 & 0.33 & 0.61 & 0.6 & 0.81 & \textbf{0.58} \\
Facelib + Facelib & 0.26 & 0.68 & 0.39 & 0.21 & 0.29 & 0.3 & 0.08 & 0.41 & 0.5 & 0.51 & 0.363\\
Detectron2 + Facelib & 0.46 & 0.65 & 0.32 & 0.53 & 0.39 & 0.69 & 0.8 & 0.54 & 0.5 & 0.76 & 0.564 \\
Detectron2-MTCNN + Facelib & 0.55 & 0.73 & 0.43 & 0.55 & 0.51 & 0.72 & 0.53 & 0.55 & 0.51 & 0.83 & \textbf{0.591} \\ 
\hline
Cascade + ConvNet & 0.34 & 0.77 & 0.45 & 0.32 & 0.32 & 0.41 & 0.12 & 0.39 & 0.52 & 0.56 & 0.42 \\
ConvNet + ConvNet & 0.27 & 0.73 & 0.46 & 0.33 & 0.26 & 0.41 & 0.12 & 0.42 & 0.51 & 0.56 & 0.40 \\
MTCNN + ConvNet & 0.52 & 0.78 & 0.59 & 0.51 & 0.56 & 0.65 & 0.43 & 0.58 & 0.57 & 0.75 & \textbf{0.594}\\
Facelib + ConvNet & 0.25 & 0.68 & 0.38 & 0.22 & 0.22 & 0.27 & 0.07 & 0.41 & 0.5 & 0.5 & 0.35\\
Detectron2 + ConvNet & 0.47 & 0.65 & 0.38 & 0.52 & 0.3 & 0.59 & 0.76 & 0.48 & 0.58 & 0.71 & 0.544 \\
Detectron2-MTCNN + ConvNet & 0.55 & 0.8 & 0.41 & 0.59 & 0.49 & 0.69 & 0.51 & 0.57 & 0.55 & 0.74 & \textbf{0.59} \\
\hline
\end{tabular}
\end{adjustwidth}
\caption{Accuracy of gender detection on the test dataset for various open source candidate systems}
\label{tab:evalresults}
\end{table}
\begin{figure}[t]
    \centering
    \includegraphics[height=6cm]{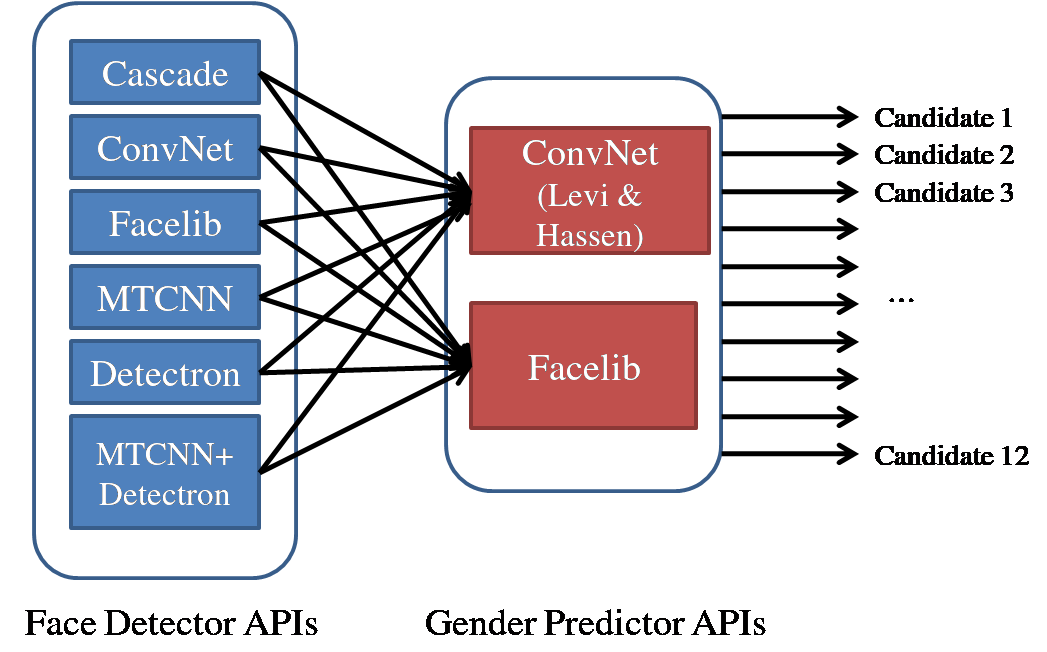}
    \caption{Combination of systems on which evaluation is carried out}
    \label{fig:combination}
\end{figure}
\subsubsection{Experimental Setup}
\label{sec:gender_setup}
We now describe our experimental setup. Our candidate frameworks are shown in Figure \ref{fig:combination}. We pick 5 different APIs for face detection; the APIs help obtain contours pertaining to faces/bodies. The images are then cropped based on the contour information with the help of the OpenCV-DNN library. We then use two different models available for gender prediction. This results in 10 different variants. Additionally, we implement a fallback mechanism for face detection, wherein if one mechanism (\textit{i.e.,} MTCNN) fails to detect faces, it will fall back to another mechanism (\textit{i.e.,} detection). Experiments are run with default configurations and evaluation is carried out on the challenge dataset described in Section \ref{sec:gender_dataset}. The source-code for this experiment is available at \url{https://github.com/swagatikadash010/gender_age.git}.

\subsubsection{Evaluation Results}
\label{sec:gender_results}
The results are shown in Table \ref{tab:evalresults}. We see that the Detectron2-mtcnn fallback mechanism for detecting faces works well with both types of gender detectors. Pipelines with MTCNN based face detectors give competitive performance and can be used running time needs to be reduced (Detectron based pipelines take around 10X longer duration to complete). It is also worth noting that for occupations where the ground truth gender distribution is imbalanced, the performance of all the variants reduces. This is expected as the chances of the number of false positives and false negatives growing are more when datasets are imbalanced. 

The confusion matrix for the best performing system is given in Table \ref{tab:confmatrix}. Looking at this matrix we can say that the model misidentifies 15 females as males which is not that satisfactory. However, for males, the model is performing better as it correctly identifies most of the images with males.

\begin{table}[t]
\centering
\begin{tabular}{ l c c c c}
\hline
& \textbf{Female} & \textbf{Male} & \textbf{Both} & \textbf{Uncertain}\\
\textbf{Female} & 23 & 15 & 4 & 2 \\
\textbf{Male} & 6 & 18 & 2 & 1 \\
\textbf{Both} & 2 & 7 & 2 & 0 \\
\textbf{Uncertain} & 3 & 2 & 1 & 13 \\
\hline
\end{tabular}
\caption{Confusion matrix for best performing system \textit{i.e.,} Detectron2-MTCNN + ConvNet}
\label{tab:confmatrix}
\end{table}
\begin{figure}[t]
    \centering
    \includegraphics[width=12cm]{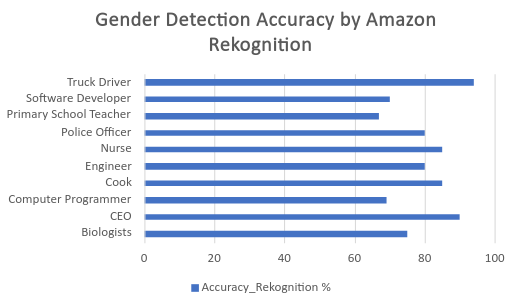}
    \caption{Accuracy scores reported by the Amazon Rekognition model for all occupations(\textbf{Avg accuracy = 79.49\%})}
    \label{fig:Accuracy reported by amazon rekognition}
\end{figure}
While open-source pipelines for gender identification are easily accessible and provide transparent and replicable outcomes on images, their performance is unreliable on open-ended images. This is because most of the pre-trained models consider frontal face data as their source of truth. One possible solution to mitigate this can be to train the same systems on large-scale open-ended images. Datasets such as Google's \textit{Open Image Dataset}\footnote{\url{https://opensource.google/projects/open-images-dataset}} can be used for this purpose. Additionally, certain architectural changes are needed to capture and aggregate additional non-facial features, such as features from the body, hands, and surrounding environment. To this end, the work by Pavlakos et al. \cite{SMPL-X:2019} is relevant, as it aims to detect demographic attributes from the positions of humans in the image, by modeling expressive body capture. 
% \subsubsection{Error Analysis and Further Improvement}
% \label{sec:gender_future}
% \begin{figure}[H]
%     \centering
%     \includegraphics[width=14cm]{images/BestOpenSource_Accuracy Scores.png}
%     \caption{Accuracy scores reported by the best open source model for all occupations(\textbf{Avg accuracy = 56.96\%})}
%     \label{fig:Accuracy reported by the best open source model}
% \end{figure}
\begin{figure}[t]
\centering
\begin{tabular}{ccc}
\subcaptionbox{Male}{\includegraphics[width = 1.2in]{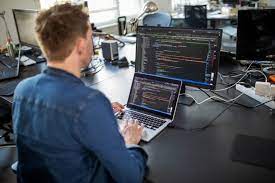}} &
\subcaptionbox{Both Male and Female}{\includegraphics[width = 1.5in]{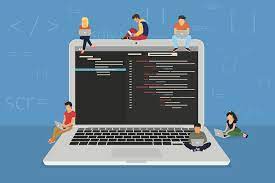}} &
\subcaptionbox{Male}{\includegraphics[width = 1.5in]{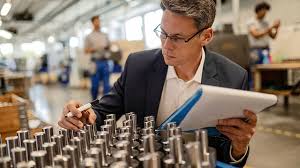}}\\
\end{tabular}
\caption{Images that are not detected by rekognition as of June 2021}
\label{fig:not_detected}
\end{figure}
\subsection{Automatic Detection of Gender of Image Search Results: Amazon Rekognition}
\label{sec:amazon_gender}
With the best open-source APIs we got the maximum accuracy of 83\% and the average accuracy score for all occupations around 60\%. We also considered one proprietary system for gender detection \textit{i.e.,} \textit{Amazon Rekognition} for detecting the gender of the faces in an image. Amazon Rekognition is based on a scalable, deep learning-based architecture for analyzing billions of images and videos with higher accuracy than its open-source counterparts. Amazon Rekognition includes a simple, easy-to-use API that can quickly analyze any image or video file that’s stored in \textit{Amazon S3}. For our project, we recorded the ``FaceDetails'' response for an image and retrieved the ``Gender'' attribute for the faces detected in the given image. With Amazon Rekognition, the average accuracy for gender detection is 79.49\% and the maximum accuracy score is for the query term ``truck driver'' which is 93.94\%. So, we considered this proprietary system for our further analysis and implementation of our de-biasing algorithm (Section \ref{sec:reranking}). However, we will also make efforts to increase the accuracy of gender detection by open-source APIs in our future work.

Some of the known issues in the Amazon Rekognition system include low accuracy in detecting read faces and non-frontal faces in general. Moreover, if an image contains faces/gender features that are overshadowed by other objects, the detection accuracy goes down. Figure \ref{fig:not_detected} shows some of such examples. 
\begin{figure}[t]
    \centering
    \includegraphics[width=12cm, height=6cm]{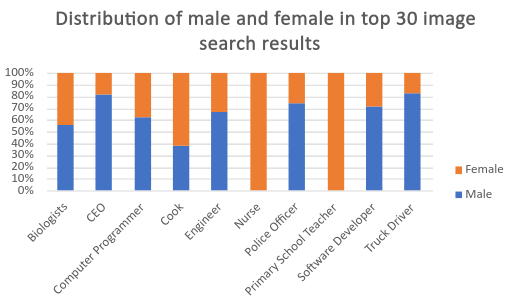}
    \caption{Distribution of male and female in top 30 Image Search Results}
    \label{fig:gender not detected by Amazon rekognition}
\end{figure}

\subsection{Insights from the distribution of gender(detected by Amazon Rekognition) across top 30 google search results}
If we consider the top 30 Google search results for all occupations and the gender distribution across these, we get the statistics presented in Figure \ref{fig:gender not detected by Amazon rekognition}. Please note that in this case, we have ignored images for which Rekognition detects ``no face'' or ``both male and female''.

From the plot, we can see that the distribution of males and females is somewhat balanced for some occupations like \textit{biologists} and \textit{computer programmers}. However, for some of the occupations like \textit{nurse} and primary school teacher, the distribution is skewed towards females to a great extent. This shows a bias in image search results for these occupations.

This sums up our exploration and implementation of frameworks for assessing occupational stereotyping in image search. While we focused on only one vital aspect of occupational stereotyping \textit{i.e.,} gender estimation, we believe that the underlying frameworks and architecture can be extended to other measures of stereotypes such as race and ethnicity. In the following section, we discuss our implementation of image re-ranking techniques that aim to produce a fairer ranked set of images for a given query (than vanilla search engine outputs), while preserving the relevance of the images with respect to the input query.
\section{\textit{Implementation:} Fairness Aware Re-Ranker}
\label{sec:reranking}
\subsection{Why re-ranking?}
Ranking reflects search engines’ estimated relevance of Web pages (or in our case, images) to the query. However, each search engine keeps its underlying ranking algorithms secret. Search engines vary by underlying ranking implementation, display of the ranked search results, and showing related pages. Proprietary systems like Google may consider many factors (some of them are personalized to a specific user) along with relevance and also they may be tuned in many ways. A study's \cite{zhao2004jump} findings suggest that a higher rank in a Google retrieval requires a combination of factors like the Google Page Rank, the popularity of websites, the density of keywords on the home page, and the keywords in the URL. However, with the secret ranking algorithm and rapid and frequent changes in these algorithms, it is impossible to have an authoritative description of these ranking algorithms. To make the search results fairer, one may argue that, the search engines and their underlying ranking algorithms can be taught to provide fairer rankings while not compromising on relevance. This is, however, a harder ask for external developers, given the opaque nature of search engines. Another way is to re-rank the retrieved results in a \textit{post-hoc} manner, to optimize fairness and relevance together. Post-hoc re-ranking not only makes the re-ranking framework agnostic to the underlying search and ranking procedure, but it also is way more scalable, transparent, and controllable. 

\subsection{Existing Work on Re-ranking}
Though systems for re-ranking of image search results remain elusive at this point, there have been several attempts to re-rank text/webpage search results. For example, the TREC 2019 and 2020 Fairness Ranking tasks \cite{biega2020overview} invite and evaluate systems for fairness ranking while maintaining the relevance for search algorithms designed to retrieve academic papers. Precisely, the goal is to provide fair exposure to different groups of authors while maintaining good relevance of the ranked papers regarding given queries. Most participating systems such as \cite{feng2020university} look to define a cost function that indicates how relevant a document is for a given query and how fair it is to a certain author group if it is ranked at a certain portion. The author groups here may correspond to the country the authors come from and the gender of the authors. For relevance, document relevance metric such as BM25 score\cite{robertson1995okapi} is considered. For fairness, Feng et al. \cite{feng2020university,feng2021towards} consider the \textit{Kullback-Leibler} (KL) divergence of the group distribution probability between the ranked list created at a certain step and the whole retrieved corpus. The rationale behind the fairness cost is this - if at any position in the final ranked list, the so-far ranked documents represent an author group distribution close to the author group distribution for the whole corpus (as measured through KL divergence), the re-ranked documents will exhibit more fairness towards the author groups. Feng et al. consider off-the-shelf systems for detecting author group attributes such as country and gender. 
%For example, if the corpus of 100 documents contains 75 male authors and 25 female authors, the top-5 re-ranked documents should roughly have 4 male and 1 female authors respectively, the top-10 re-ranked documents should roughly have 8 male and 2 female authors, and so on. 
We implement this strategy for re-ranking of image results, which we describe in the following sections. 
\begin{figure}[t]
    \centering
    \includegraphics[width=12cm]{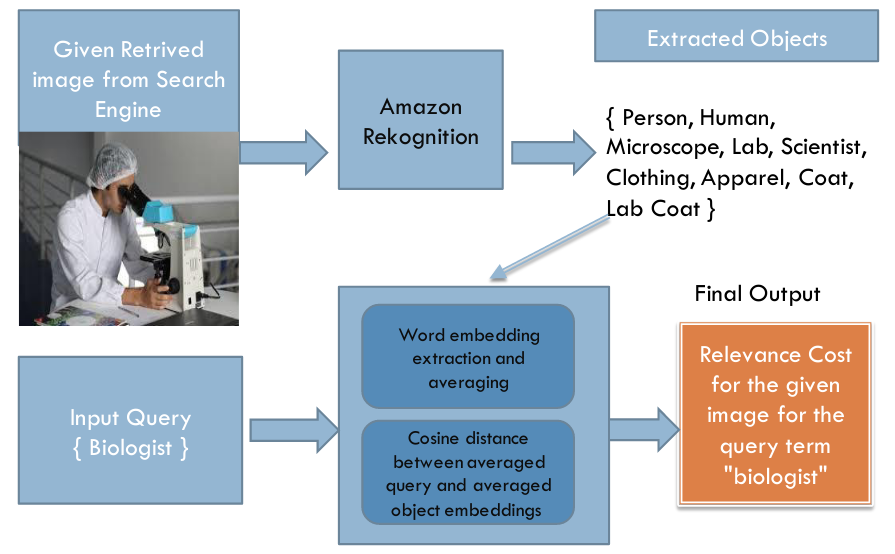}
    \caption{A pictorial representation of measuring relevance score}
    \label{fig:Measurement_Relevance}
\end{figure}
\subsection{Methodology}
We propose a fairness ranking algorithm for images incorporating both relevance and fairness with respect to the gender distribution for the given search query. The objective is to assign a higher rank to an image from a lot that maximizes a defined relevance score (or equivalently minimizes the relevance cost) while ensuring fairness. For relevance cost measurement, we propose a scheme depicted in Figure \ref{fig:Measurement_Relevance}. For a given query (say ``biologist'') and a retrieved image, we first extract a set of object labels from the image. We use an off-the-shelf system such as Amazon Rekognition (``detect\_label'' handle) for object detection. Once, object terms are identified, we extract their word embedding representation using GloVe embeddings \cite{pennington2014glove}. We use the ``glove-wiki-gigaword-300-binary'' pretrained model.  GloVe is an unsupervised learning algorithm for obtaining vector representations for words.  Training of GloVe is performed on aggregated global word-to-word co-occurrence statistics from a large corpus, and the resulting representations showcase interesting linear substructures of the word vector space. Once embeddings are extracted, they are averaged and we compute the cosine distance between the averaged object-term vector and the vector that represents the query word. Intuitively, the distance indicates how dissimilar the set of objects is with the query. The more the dissimilarity, more becomes the relevance cost. 

Our re-ranking method is given as follows. Let us assume that for a given query $q$ a set of images $I'$ are already retrieved using an image search engine, from a large indexed image corpus $I$. Our intention is to re-rank $I'$ and form a re-ranked list $R$. We initialize $R$ with an empty list and gradually move an image $i$ from $I'$. At any given time, the idea is to select an image $i$ in such a way that it minimizes the overall cost of adding it to $R$. The overall cost is given below:
\begin{equation}
\begin{split}
C(i, w, R, I', q) = w_r * cosine\_dist(w2v([objects_i]),w2v([q])) \\
+ w_g *KL(p(g,R+{i}) || p(g, I'))
\end{split}
\label{eq:cost}
\end{equation}

where, $w_r$ and $w_g$ are user-defined weights, that control how much each term in the above equation, contributes to the overall cost of adding an image ($w_r+w_g = 1$). $w2v(.)$ and $cosine\_dist(.)$ are functions to compute average word embeddings of given terms and cosine distance between two embeddings (vectors). $objects_i$ represents objects identified from image $i$. $p(g, .)$ represents the distribution of group property (in our case, gender) in a given set of observations. $KL(.)$ is the \textit{Kullback-Leibler} divergence between two distributions. The rationale behind this formulation is similar to that of Feng et al. \cite{feng2020university,feng2021towards}.
\subsection{Experimental Setup}
We now describe our experimental setup, datasets and implementation and evaluation details. The implementation can be checked out from \url{https://github.com/swagatikadash010/Image_Search_Bias}.
\subsubsection{Dataset}
We use the same dataset described in Section \ref{sec:gender_dataset}. We have top 100 Google search results for each of the 10 occupations i.e, ``Biologist'', ``CEO'', ``Cook'', ``Engineer'', ``Nurse'', ``Police Officer'', ``Primary School Teacher'', ``Programmer'', ``Software developer'', ``Truck driver''. The original rankings given by Google is used as ground truth for relevance scores, which will serve as a reference for relevance metric computation. Additionally, we have ground-truth files mentioning the gender of the persons present in the images, obtained through crowd sourcing. This will help in evaluating the fairness metric.
\subsubsection{Baselines and Systems for Comparison}
We used random ranking and only relevance score based ranking as our baselines for comparison. We vary the $w_r$ and $w_g$ weights given in Equation \ref{eq:cost} and experiment with $w_r = [0.1,0.3,0.5,0.7,0.9]$ and $w_g = (1-w_r)$. It is worth noting that, unlike Feng et al. \cite{feng2021towards}, we do not normalize the relevance cost as cosine distance is bounded between 0 and 1.0 for word embeddings, in practice.
\begin{figure}[t]
    \centering
    \includegraphics[width=10cm]{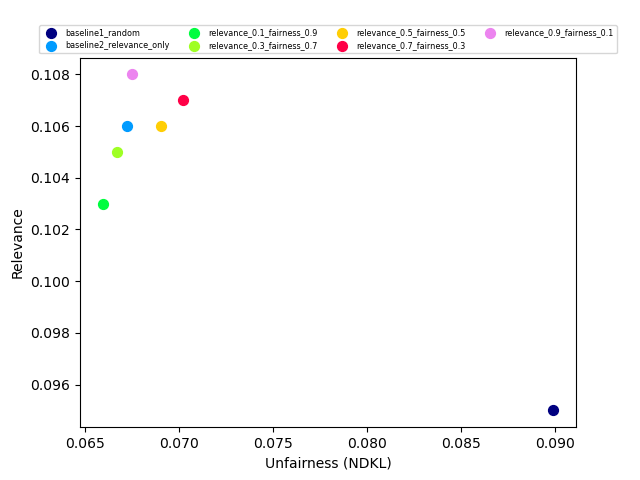}
    \caption{Performance plot for different models averaged over 10 occupations}
    \label{fig:PerformancePlot_Overall}
\end{figure}
\begin{figure}[t]
\begin{adjustwidth}{-1.2in}{-0.1in}
\captionsetup[subfigure]{labelformat=empty}
    \centering % <-- added
\begin{subfigure}{0.50\textwidth}
\subfloat[CEO]{%
  \includegraphics[width=8cm, height = 8cm]{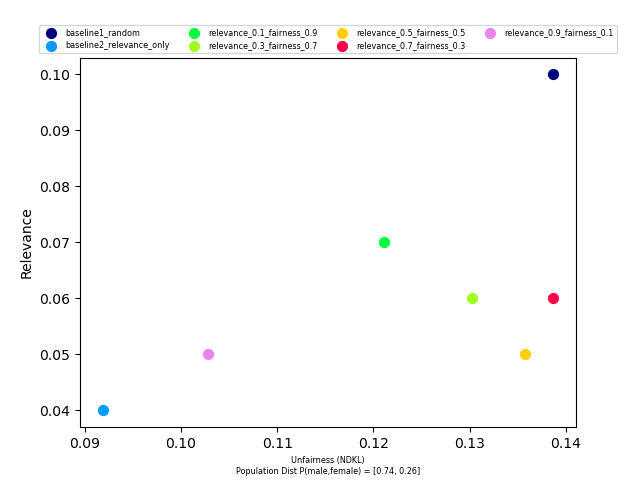}}
  \label{fig:subfig_ceo}
\end{subfigure}\hfil
\begin{subfigure}{0.40\textwidth}
\subfloat[Nurse]{%
  \includegraphics[width=8cm, height = 8cm]{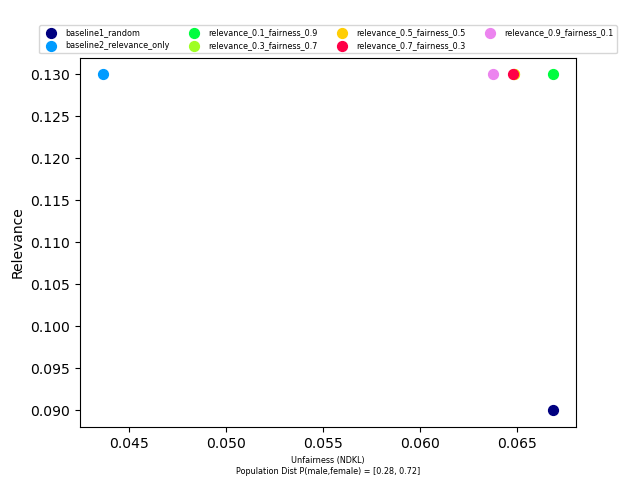}}
  \label{fig:subfig_nurse}
\end{subfigure}\hfil
\end{adjustwidth}
\caption{Performance plot for queries ``CEO'' and ``Nurse''}
\label{fig:twoplots}
\end{figure}
\subsubsection{Evaluation Criteria}
We consider two metrics for evaluating the systems: (a) Relevance and (b) Fairness. For relevance, we consider \textit{bucket ranking accuracy} of the systems. This is because, unlike document search, the absolute ranking does not make much sense as image search results are often displayed in a grid-like structure and are not shown in a list-like structure as documents. We surmise that images in a certain bucket should be equally relevant. For example, if we consider that the first 30 images are visible to the user on one page and the next 30 appear on the second page and so on, images in the first bucket of size 30 are equally relevant. Based on this idea, we map the ground truth relevance scores (Google's ranks) and predicted ranks into buckets of 30, and compute the relevance score as follows:
\begin{equation}
Relevance = \frac{\#(predicted\_rank==ground\_truth\_rank)}{\#total\_images}
\label{eq:relevance}
\end{equation}

For fairness, we follow Geyik et al. \cite{geyik2019fairness} and compute the Normalized Discounted KL Divergence (NDKL) as a measure of the degree of unfairness. This is given as: 
\begin{equation}
    Unfairness (R) = \frac{1}{Z} \sum_{i=1}^{|R|}\frac{1}{log_{2}(i+1))} * KL(p(g,R_{i})~||~p(g,D'))
\label{eq:unfairness}
\end{equation}

where, $$ Z = \sum_{i=1}^{|R|}\frac{1}{log_{2}(i+1))}$$.
%=================
\begin{figure}[t]
\captionsetup[subfigure]{labelformat=empty}
    %\centering
\begin{subfigure}{0.20\textwidth}
\subfloat[Rank 1]{%
  \includegraphics[width=0.9\linewidth, height = 2 cm]{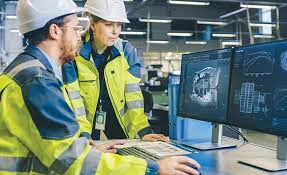}}
  %\caption{Rank 4}
  \label{fig:subfig_anecdote1}
\end{subfigure}\hfil
\begin{subfigure}{0.20\textwidth}
\subfloat[Rank 2]{%
  \includegraphics[width=0.9\linewidth, height = 2 cm]{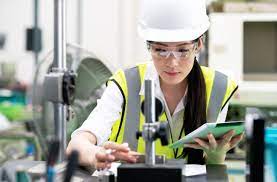}}
  %\caption{Rank 4}
  \label{fig:subfig_anecdote2}
\end{subfigure}\hfil
\begin{subfigure}{0.20\textwidth}
\subfloat[Rank 3]{%
  \includegraphics[width=0.9\linewidth, height = 2 cm]{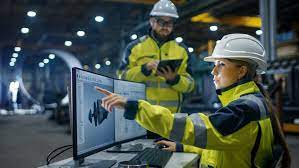}}
  %\caption{Rank 4}
  \label{fig:subfig_anecdote3}
\end{subfigure}\hfil
\begin{subfigure}{0.20\textwidth}
\subfloat[Rank 4]{%
  \includegraphics[width=0.9\linewidth, height = 2 cm]{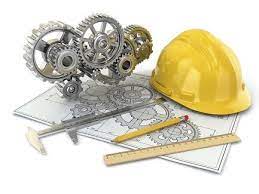}}
  %\caption{Rank 4}
  \label{fig:subfig_anecdote4}
\end{subfigure}\hfil
\begin{subfigure}{0.20\textwidth}
\subfloat[Rank 5]{%
  \includegraphics[width=0.9\linewidth, height = 2 cm]{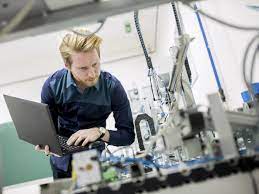}}
  %\caption{Rank 4}
\end{subfigure}\hfil
\medskip
\begin{subfigure}{0.20\textwidth}
\subfloat[Rank 1]{%
  \includegraphics[width=0.9\linewidth, height = 2 cm]{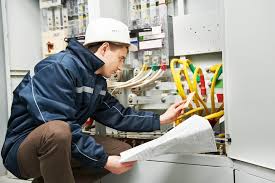}}
  %\caption{Rank 4}
  \label{fig:subfig_anecdote6}
\end{subfigure}\hfil
\begin{subfigure}{0.20\textwidth}
\subfloat[Rank 2]{%
  \includegraphics[width=0.9\linewidth, height = 2 cm]{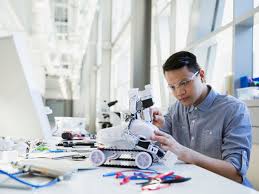}}
  %\caption{Rank 4}
  \label{fig:subfig_anecdote7}
\end{subfigure}\hfil
\begin{subfigure}{0.20\textwidth}
\subfloat[Rank 3]{%
  \includegraphics[width=0.9\linewidth, height = 2 cm]{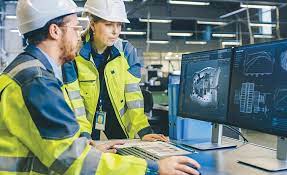}}
  %\caption{Rank 4}
  \label{fig:subfig_anecdote8}
\end{subfigure}\hfil
\begin{subfigure}{0.20\textwidth}
\subfloat[Rank 4]{%
  \includegraphics[width=0.9\linewidth, height = 2 cm]{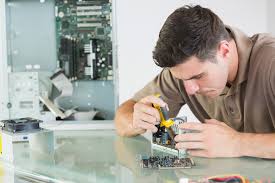}}
  %\caption{Rank 4}
  \label{fig:subfig_anecdote9}
\end{subfigure}\hfil
\begin{subfigure}{0.20\textwidth}
\subfloat[Rank 5]{%
  \includegraphics[width=0.9\linewidth, height = 2 cm]{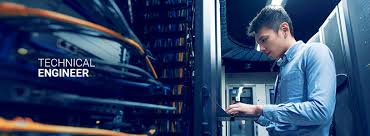}}
  %\caption{Rank 4}
  \label{fig:subfig_anecdote10}
\end{subfigure}\hfil
\medskip
\begin{subfigure}{0.20\textwidth}
\subfloat[0, Male]{%
  \includegraphics[width=0.9\linewidth, height = 2 cm]{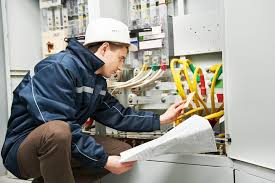}}
  %\caption{Rank 4}
  \label{fig:subfig_anecdote11}
\end{subfigure}\hfil
\begin{subfigure}{0.20\textwidth}
\subfloat[1, Male]{%
  \includegraphics[width=0.9\linewidth, height = 2 cm]{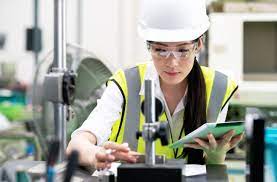}}
  %\caption{Rank 4}
  \label{fig:subfig_anecdote12}
\end{subfigure}\hfil
\begin{subfigure}{0.20\textwidth}
\subfloat[2, Female]{%
  \includegraphics[width=0.9\linewidth, height = 2 cm]{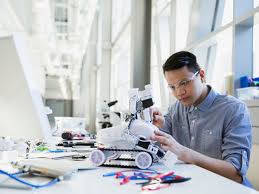}}
  %\caption{Rank 4}
  \label{fig:subfig_anecdote13}
\end{subfigure}\hfil
\begin{subfigure}{0.20\textwidth}
\subfloat[3, Male]{%
  \includegraphics[width=0.9\linewidth, height = 2 cm]{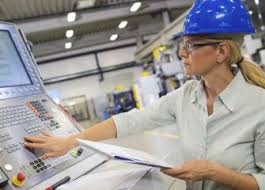}}
  %\caption{Rank 4}
  \label{fig:subfig_anecdote14}
\end{subfigure}\hfil
\begin{subfigure}{0.20\textwidth}
\subfloat[4, Male]{%
  \includegraphics[width=0.9\linewidth, height = 2 cm]{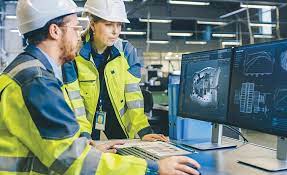}}
  %\caption{Rank 4}
  \label{fig:subfig_anecdote15}
\end{subfigure}\hfil
\caption{Example rankings given by Google (Row 1), relevance only (Row 2) and  weighted relevance and fairness model (Row 3) for Keyword ``Engineer''. Search was conducted in June 2021.}
\label{fig:fairnessranks}
\end{figure}
Intuitively, this metric penalizes the degree of unfairness (computed through KL divergence) observed in higher-ranked documents and discounts unfairness as we proceed through the rankings.
\subsection{Results and Analysis}
The overall results are plotted in Figure \ref{fig:PerformancePlot_Overall}. As expected, the random baseline does not re-rank very well and a trade-off between relevance and fairness is seen. There is no clear winner which optimizes both aspects the best and models can be selected based on how sensitive these two aspects can become for specific search applications. 

While, for most of the occupations, we get similar plots as the one shown above, queries like ``CEO'' and ``Nurse'' yielded different observations. Figure \ref{fig:twoplots} presents performance plots for these two keywords. For ``CEO'', the overall relevance score is very low even for the relevance-only baseline. This is because, for CEO the extracted labels are not semantically much relevant to the keyword ``Chief Executive Officer''. The minimum relevance score is 0.6974 for all 100 images. For ``Nurse'' the gender distribution of the retrieved corpus is very highly biased towards Females (around 72\%). Biases of such kind, may not have helped the KL divergence term to yield meaningful indications of fairness cost.

Figure \ref{fig:fairnessranks} presents some anecdotal examples where we qualitatively compare Google-ranked images, images ranked by relevance-only baseline, and by our fairness-aware algorithm. The query keyword here is ``Engineer''. As expected we see that while the relevance-only model provides images that are quite relevant for the query term engineer, they are quite male-dominated. This is mitigated by our fairness-aware algorithm (here we have set $w_r = w_g= 0.5$) which distributes the images better across different genders. 
\section{Conclusion and Future Directions}
\label{sec:conclusion}
Search and retrieval systems have become an integral part of human lives and have attained remarkable success and the trust of users in recent times. Their efficacy in fairly retrieving and representing, however, remains below par and this has raised concerns in the information retrieval community. In this paper, we discussed our explorations on the fairness of ranking in search engines, primarily focusing on gender stereotyping issues in occupational keyword-based image search. We discussed the fairness issues that arise from default search and retrieval mechanisms and proposed a fairness-aware ranking procedure that can help mitigate the bias. For gender bias assessment, we employed both open-source pre-trained models and a proprietary system like Amazon Rekognition for gender identification. This helped us in gauging the gender bias in search results obtained for several occupational keywords. For de-biasing, our proposed ranking algorithm uses the gender identification APIs and models and re-ranks the retrieved images through a carefully designed cost function that considers both relevance and fairness. On received sets of images for 10 occupational keywords,  we plotted the performance of our de-biased model and compared it with the baseline systems having random and relevance-only re-ranking methods. Our experimental results justified our proposed model performing better in terms of fairness of the image search results. 
\subsection{Future Work}
\begin{itemize}
    \item The maximum average accuracy for gender detection for the open-source and the proprietary system is 59.1\% and 79.49\% respectively. This lower accuracy is primarily due to the open-ended nature of the images and the absence of prominent facial features. In the future, we will explore models that consider the body and environmental features to classify genders better.
    \item For measuring relevance, we extracted the labels using Amazon Rekognition API, and using word embedding we calculated the semantic similarity between the extracted labels with the occupation keyword. In the future, we will consider joint language and vision models like VisualBERT\cite{li2019visualbert} to compute the relevance scores.
    \item For de-biasing, we only considered a cost-based re-ranking algorithm that does not improve the search over time. We can use this cost to optimize the ranking procedure itself with the help of reinforcement learning. 
    \item This study only considered the search results from Google. We will take image search results from other popular search engines and open-source search frameworks for our experiments.
    \item This study only included occupational stereotypes and it is aligned with a few types of biases(as described in the section \ref{sec:type_biases}). Mitigating other forms of biases (such as racial bias) is also on our agenda.
\end{itemize}

\section*{Acknowledgements}
We would like to thank Professor Yunhe Feng, Department of Computer Science and Engineering, University of North Texas and Professor Chirag Shah, School of Information, University of Washington, for their continuous guidance and support.

\bibliography{bibliography}

\begin{thebibliography}{59}
\providecommand{\natexlab}[1]{#1}
\providecommand{\url}[1]{\texttt{#1}}
\expandafter\ifx\csname urlstyle\endcsname\relax
  \providecommand{\doi}[1]{doi: #1}\else
  \providecommand{\doi}{doi: \begingroup \urlstyle{rm}\Url}\fi

\bibitem[Biega et~al.(2020)Biega, Diaz, Ekstrand, and
  Kohlmeier]{biega2020overview}
Asia~J Biega, Fernando Diaz, Michael~D Ekstrand, and Sebastian Kohlmeier.
\newblock Overview of the trec 2019 fair ranking track.
\newblock \emph{arXiv preprint arXiv:2003.11650}, 2020.

\bibitem[Bond et~al.(2012)Bond, Fariss, Jones, Kramer, Marlow, Settle, and
  Fowler]{bond201261}
Robert~M Bond, Christopher~J Fariss, Jason~J Jones, Adam~DI Kramer, Cameron
  Marlow, Jaime~E Settle, and James~H Fowler.
\newblock A 61-million-person experiment in social influence and political
  mobilization.
\newblock \emph{Nature}, 489\penalty0 (7415):\penalty0 295--298, 2012.

\bibitem[Broder(2002)]{broder2002taxonomy}
Andrei Broder.
\newblock A taxonomy of web search.
\newblock In \emph{ACM Sigir forum}, volume~36, pp.\  3--10. ACM New York, NY,
  USA, 2002.

\bibitem[Chakraborty et~al.(2000)Chakraborty, Kaustubha, Hegde, Pereira, Done,
  Kirlin, Moghaddamjoo, Georgakis, Kotropoulos, and
  Xafopoulos]{chakraborty2000bishop}
B~Chakraborty, R~Kaustubha, A~Hegde, A~Pereira, W~Done, R~Kirlin,
  A~Moghaddamjoo, A~Georgakis, C~Kotropoulos, and Pitas Xafopoulos.
\newblock Bishop, cm, neural networks for pattern recognition, oxford
  university press, new york, 1995. carreira-perpi{\~n}{\'a}n m., mode-finding
  for mixtures of gaussian distributions, ieee transaction on pattern analysis
  and machine intelligence, vol. 22, no. 11, november 2000, 1318-1323.
\newblock \emph{IEEE transaction on Pattern Analysis and Machine Intelligence},
  22\penalty0 (11):\penalty0 1318--1323, 2000.

\bibitem[Chernov(2019)]{Ageandge68:online}
Pavel Chernov.
\newblock Age and gender estimation. open-source projects overview. simple
  project from scratch.
\newblock
  https://medium.com/@pavelchernov/age-and-gender-estimation-open-source-projects-overview-simple-project-from-scratch-69581831297e,
  2019.
\newblock (Accessed on 04/28/2023).

\bibitem[Craswell et~al.(2008)Craswell, Zoeter, Taylor, and
  Ramsey]{craswell2008experimental}
Nick Craswell, Onno Zoeter, Michael Taylor, and Bill Ramsey.
\newblock An experimental comparison of click position-bias models.
\newblock In \emph{Proceedings of the 2008 international conference on web
  search and data mining}, pp.\  87--94, 2008.

\bibitem[{\'C}urkovi{\'c} \& Ko{\v{s}}ec(2018){\'C}urkovi{\'c} and
  Ko{\v{s}}ec]{curkovic2018bubble}
Marko {\'C}urkovi{\'c} and Andro Ko{\v{s}}ec.
\newblock Bubble effect: including internet search engines in systematic
  reviews introduces selection bias and impedes scientific reproducibility.
\newblock \emph{BMC medical research methodology}, 18\penalty0 (1):\penalty0
  1--3, 2018.

\bibitem[Deng et~al.(2009)Deng, Dong, Socher, Li, Li, and
  Fei-Fei]{deng2009imagenet}
Jia Deng, Wei Dong, Richard Socher, Li-Jia Li, Kai Li, and Li~Fei-Fei.
\newblock Imagenet: A large-scale hierarchical image database.
\newblock In \emph{2009 IEEE conference on computer vision and pattern
  recognition}, pp.\  248--255. Ieee, 2009.

\bibitem[Dutton et~al.(2013)Dutton, Blank, and Groselj]{dutton2013cultures}
William~H Dutton, Grant Blank, and Darja Groselj.
\newblock \emph{Cultures of the internet: the internet in Britain: Oxford
  Internet Survey 2013 Report}.
\newblock Oxford Internet Institute, 2013.

\bibitem[Dutton et~al.(2017)Dutton, Reisdorf, Dubois, and
  Blank]{dutton2017search}
William~H Dutton, Bianca Reisdorf, Elizabeth Dubois, and Grant Blank.
\newblock \emph{Search and politics: The uses and impacts of search in Britain,
  France, Germany, Italy, Poland, Spain, and the United States}.
\newblock Quello Center Working Paper, 2017.

\bibitem[Epstein \& Robertson(2015)Epstein and Robertson]{epstein2015search}
Robert Epstein and Ronald~E Robertson.
\newblock The search engine manipulation effect (seme) and its possible impact
  on the outcomes of elections.
\newblock \emph{Proceedings of the National Academy of Sciences}, 112\penalty0
  (33):\penalty0 E4512--E4521, 2015.

\bibitem[Epstein et~al.(2017)Epstein, Robertson, Lazer, and
  Wilson]{epstein2017suppressing}
Robert Epstein, Ronald~E Robertson, David Lazer, and Christo Wilson.
\newblock Suppressing the search engine manipulation effect (seme).
\newblock \emph{Proceedings of the ACM on Human-Computer Interaction},
  1\penalty0 (CSCW):\penalty0 1--22, 2017.

\bibitem[Feng et~al.(2020)Feng, Saelid, Li, Gao, and Shah]{feng2020university}
Yunhe Feng, Daniel Saelid, Ke~Li, Ruoyuan Gao, and Chirag Shah.
\newblock University of washington at trec 2020 fairness ranking track.
\newblock \emph{arXiv preprint arXiv:2011.02066}, 2020.

\bibitem[Feng et~al.(2021)Feng, Saelid, Li, Gao, and Shah]{feng2021towards}
Yunhe Feng, Daniel Saelid, Ke~Li, Ruoyuan Gao, and Chirag Shah.
\newblock Towards fairness-aware ranking by defining latent groups using
  inferred features.
\newblock In \emph{International Workshop on Algorithmic Bias in Search and
  Recommendation}, pp.\  1--8. Springer, 2021.

\bibitem[Fidel(2012)]{fidel2012human}
Raya Fidel.
\newblock \emph{Human information interaction: An ecological approach to
  information behavior}.
\newblock Mit Press, 2012.

\bibitem[Fogg(2002)]{fogg2002persuasive}
Brian~J Fogg.
\newblock Persuasive technology: using computers to change what we think and
  do.
\newblock \emph{Ubiquity}, 2002\penalty0 (December):\penalty0 2, 2002.

\bibitem[Geyik et~al.(2019)Geyik, Ambler, and Kenthapadi]{geyik2019fairness}
Sahin~Cem Geyik, Stuart Ambler, and Krishnaram Kenthapadi.
\newblock Fairness-aware ranking in search \& recommendation systems with
  application to linkedin talent search.
\newblock In \emph{Proceedings of the 25th acm sigkdd international conference
  on knowledge discovery \& data mining}, pp.\  2221--2231, 2019.

\bibitem[Gillespie(2014)]{gillespie2014relevance}
Tarleton Gillespie.
\newblock The relevance of algorithms.
\newblock \emph{Media technologies: Essays on communication, materiality, and
  society}, 167\penalty0 (2014):\penalty0 167, 2014.

\bibitem[Grind et~al.(2019)Grind, Schechner, McMillan, and
  West]{grind2019google}
Kirsten Grind, Sam Schechner, Robert McMillan, and John West.
\newblock How google interferes with its search algorithms and changes your
  results.
\newblock \emph{The Wall Street Journal}, 15, 2019.

\bibitem[Guarino(2016)]{guarino2016google}
Ben Guarino.
\newblock Google faulted for racial bias in image search results for black
  teenagers.
\newblock \emph{Washington Post}, 6:\penalty0 2016, 2016.

\bibitem[Hardt et~al.(2016)Hardt, Price, and Srebro]{hardt2016equality}
Moritz Hardt, Eric Price, and Nati Srebro.
\newblock Equality of opportunity in supervised learning.
\newblock \emph{Advances in neural information processing systems},
  29:\penalty0 3315--3323, 2016.

\bibitem[Hopkins et~al.(2021)Hopkins, O'Neil, Bilimoria, and
  Broadfoot]{hopkins2021buried}
Margaret~M Hopkins, Deborah~Anne O'Neil, Diana Bilimoria, and Alison Broadfoot.
\newblock Buried treasure: Contradictions in the perception and reality of
  women's leadership.
\newblock \emph{Frontiers in Psychology}, 12:\penalty0 1804, 2021.

\bibitem[Howard et~al.(2017)Howard, Zhu, Chen, Kalenichenko, Wang, Weyand,
  Andreetto, and Adam]{howard2017mobilenets}
Andrew~G Howard, Menglong Zhu, Bo~Chen, Dmitry Kalenichenko, Weijun Wang,
  Tobias Weyand, Marco Andreetto, and Hartwig Adam.
\newblock Mobilenets: Efficient convolutional neural networks for mobile vision
  applications.
\newblock \emph{arXiv preprint arXiv:1704.04861}, 2017.

\bibitem[Jain \& Varma(2011)Jain and Varma]{jain2011learning}
Vidit Jain and Manik Varma.
\newblock Learning to re-rank: query-dependent image re-ranking using click
  data.
\newblock In \emph{Proceedings of the 20th international conference on World
  wide web}, pp.\  277--286, 2011.

\bibitem[Joachims et~al.(2007)Joachims, Granka, Pan, Hembrooke, Radlinski, and
  Gay]{joachims2007evaluating}
Thorsten Joachims, Laura Granka, Bing Pan, Helene Hembrooke, Filip Radlinski,
  and Geri Gay.
\newblock Evaluating the accuracy of implicit feedback from clicks and query
  reformulations in web search.
\newblock \emph{ACM Transactions on Information Systems (TOIS)}, 25\penalty0
  (2):\penalty0 7--es, 2007.

\bibitem[Kay et~al.(2015)Kay, Matuszek, and Munson]{kay2015unequal}
Matthew Kay, Cynthia Matuszek, and Sean~A Munson.
\newblock Unequal representation and gender stereotypes in image search results
  for occupations.
\newblock In \emph{Proceedings of the 33rd Annual ACM Conference on Human
  Factors in Computing Systems}, pp.\  3819--3828, 2015.

\bibitem[Kilbertus et~al.(2017)Kilbertus, Rojas-Carulla, Parascandolo, Hardt,
  Janzing, and Sch{\"o}lkopf]{kilbertus2017avoiding}
Niki Kilbertus, Mateo Rojas-Carulla, Giambattista Parascandolo, Moritz Hardt,
  Dominik Janzing, and Bernhard Sch{\"o}lkopf.
\newblock Avoiding discrimination through causal reasoning.
\newblock \emph{arXiv preprint arXiv:1706.02744}, 2017.

\bibitem[Kulshrestha et~al.(2017)Kulshrestha, Eslami, Messias, Zafar, Ghosh,
  Gummadi, and Karahalios]{kulshrestha2017quantifying}
Juhi Kulshrestha, Motahhare Eslami, Johnnatan Messias, Muhammad~Bilal Zafar,
  Saptarshi Ghosh, Krishna~P Gummadi, and Karrie Karahalios.
\newblock Quantifying search bias: Investigating sources of bias for political
  searches in social media.
\newblock In \emph{Proceedings of the 2017 ACM Conference on Computer Supported
  Cooperative Work and Social Computing}, pp.\  417--432, 2017.

\bibitem[Langston(2015)]{langston2015sa}
Jennifer Langston.
\newblock Who’sa ceo? google image results can shift gender biases.
\newblock \emph{UW News, April}, 2015.

\bibitem[Levi \& Hassner(2015)Levi and Hassner]{LH:CVPRw15:age}
Gil Levi and Tal Hassner.
\newblock Age and gender classification using convolutional neural networks.
\newblock In \emph{IEEE Conf. on Computer Vision and Pattern Recognition (CVPR)
  workshops}, 2015.
\newblock URL
  \url{https://osnathassner.github.io/talhassner/projects/cnn_agegender}.

\bibitem[Li et~al.(2019)Li, Yatskar, Yin, Hsieh, and Chang]{li2019visualbert}
Liunian~Harold Li, Mark Yatskar, Da~Yin, Cho-Jui Hsieh, and Kai-Wei Chang.
\newblock Visualbert: A simple and performant baseline for vision and language.
\newblock \emph{arXiv preprint arXiv:1908.03557}, 2019.

\bibitem[Lim et~al.(2020)Lim, Jatowt, F{\"a}rber, and
  Yoshikawa]{lim-etal-2020-annotating}
Sora Lim, Adam Jatowt, Michael F{\"a}rber, and Masatoshi Yoshikawa.
\newblock Annotating and analyzing biased sentences in news articles using
  crowdsourcing.
\newblock In \emph{Proceedings of the 12th Language Resources and Evaluation
  Conference}, pp.\  1478--1484, Marseille, France, May 2020. European Language
  Resources Association.
\newblock ISBN 979-10-95546-34-4.
\newblock URL \url{https://aclanthology.org/2020.lrec-1.184}.

\bibitem[Liu(2011)]{liu2011learning}
Tie-Yan Liu.
\newblock \emph{Learning to rank for information retrieval}.
\newblock Springer Science \& Business Media, 2011.

\bibitem[Mai(2016)]{mai2016looking}
Jens-Erik Mai.
\newblock \emph{Looking for information: A survey of research on information
  seeking, needs, and behavior}.
\newblock Emerald Group Publishing, 2016.

\bibitem[Mitchell et~al.(2017)Mitchell, Gottfried, Shearer, and
  Lu]{mitchell2017americans}
Amy Mitchell, Jeffrey Gottfried, Elisa Shearer, and Kristine Lu.
\newblock \emph{How Americans encounter, recall and act upon digital news}.
\newblock Pew Research Center, 2017.

\bibitem[Pasquale(2015)]{pasquale2015black}
Frank Pasquale.
\newblock \emph{The black box society}.
\newblock Harvard University Press, 2015.

\bibitem[Pavlakos et~al.(2019)Pavlakos, Choutas, Ghorbani, Bolkart, Osman,
  Tzionas, and Black]{SMPL-X:2019}
Georgios Pavlakos, Vasileios Choutas, Nima Ghorbani, Timo Bolkart, Ahmed A.~A.
  Osman, Dimitrios Tzionas, and Michael~J. Black.
\newblock Expressive body capture: 3d hands, face, and body from a single
  image.
\newblock In \emph{Proceedings IEEE Conf. on Computer Vision and Pattern
  Recognition (CVPR)}, 2019.

\bibitem[Pennington et~al.(2014)Pennington, Socher, and
  Manning]{pennington2014glove}
Jeffrey Pennington, Richard Socher, and Christopher~D Manning.
\newblock Glove: Global vectors for word representation.
\newblock In \emph{Proceedings of the 2014 conference on empirical methods in
  natural language processing (EMNLP)}, pp.\  1532--1543, 2014.

\bibitem[Robertson et~al.(2018)Robertson, Jiang, Joseph, Friedland, Lazer, and
  Wilson]{robertson2018auditing}
Ronald~E Robertson, Shan Jiang, Kenneth Joseph, Lisa Friedland, David Lazer,
  and Christo Wilson.
\newblock Auditing partisan audience bias within google search.
\newblock \emph{Proceedings of the ACM on Human-Computer Interaction},
  2\penalty0 (CSCW):\penalty0 1--22, 2018.

\bibitem[Robertson et~al.(1995)Robertson, Walker, Jones, Hancock-Beaulieu,
  Gatford, et~al.]{robertson1995okapi}
Stephen~E Robertson, Steve Walker, Susan Jones, Micheline~M Hancock-Beaulieu,
  Mike Gatford, et~al.
\newblock Okapi at trec-3.
\newblock \emph{Nist Special Publication Sp}, 109:\penalty0 109, 1995.

\bibitem[Rocchio(1971)]{rocchio1971smart}
Joseph~John Rocchio.
\newblock The smart retrieval system: Experiments in automatic document
  processing.
\newblock \emph{Relevance feedback in information retrieval}, pp.\  313--323,
  1971.

\bibitem[Rothe et~al.(2015)Rothe, Timofte, and Van~Gool]{rothe2015dex}
Rasmus Rothe, Radu Timofte, and Luc Van~Gool.
\newblock Dex: Deep expectation of apparent age from a single image.
\newblock In \emph{Proceedings of the IEEE international conference on computer
  vision workshops}, pp.\  10--15, 2015.

\bibitem[Sandvig et~al.(2014)Sandvig, Hamilton, Karahalios, and
  Langbort]{sandvig2014auditing}
Christian Sandvig, Kevin Hamilton, Karrie Karahalios, and Cedric Langbort.
\newblock Auditing algorithms: Research methods for detecting discrimination on
  internet platforms.
\newblock \emph{Data and discrimination: converting critical concerns into
  productive inquiry}, 22:\penalty0 4349--4357, 2014.

\bibitem[Schroeder \& Borgerson(2015)Schroeder and
  Borgerson]{schroeder2015critical}
Jonathan~E Schroeder and Janet~L Borgerson.
\newblock Critical visual analysis of gender: Reactions and reflections.
\newblock \emph{Journal of Marketing Management}, 31\penalty0 (15-16):\penalty0
  1723--1731, 2015.

\bibitem[Schweiger et~al.(2014)Schweiger, Oeberst, and
  Cress]{schweiger2014confirmation}
Stefan Schweiger, Aileen Oeberst, and Ulrike Cress.
\newblock Confirmation bias in web-based search: a randomized online study on
  the effects of expert information and social tags on information search and
  evaluation.
\newblock \emph{Journal of medical Internet research}, 16\penalty0
  (3):\penalty0 e94, 2014.

\bibitem[Silberg \& Manyika(2019)Silberg and Manyika]{silberg2019notes}
Jake Silberg and James Manyika.
\newblock Notes from the ai frontier: Tackling bias in ai (and in humans).
\newblock \emph{McKinsey Global Institute (June 2019)}, 2019.

\bibitem[Sutton \& Barto(2018)Sutton and Barto]{sutton2018reinforcement}
Richard~S Sutton and Andrew~G Barto.
\newblock \emph{Reinforcement learning: An introduction}.
\newblock MIT press, 2018.

\bibitem[Viola \& Jones(2001)Viola and Jones]{viola2001rapid}
Paul Viola and Michael Jones.
\newblock Rapid object detection using a boosted cascade of simple features.
\newblock In \emph{Proceedings of the 2001 IEEE computer society conference on
  computer vision and pattern recognition. CVPR 2001}, volume~1, pp.\  I--I.
  Ieee, 2001.

\bibitem[Wu et~al.(2019)Wu, Kirillov, Massa, Lo, and
  Girshick]{wu2019detectron2}
Yuxin Wu, Alexander Kirillov, Francisco Massa, Wan-Yen Lo, and Ross Girshick.
\newblock Detectron2.
\newblock \url{https://github.com/facebookresearch/detectron2}, 2019.

\bibitem[Xie et~al.(2019)Xie, Mao, Liu, de~Rijke, Shao, Ye, Zhang, and
  Ma]{xie2019grid}
Xiaohui Xie, Jiaxin Mao, Yiqun Liu, Maarten de~Rijke, Yunqiu Shao, Zixin Ye,
  Min Zhang, and Shaoping Ma.
\newblock Grid-based evaluation metrics for web image search.
\newblock In \emph{The World Wide Web Conference}, pp.\  2103--2114, 2019.

\bibitem[Yang et~al.(2018)Yang, Huang, Lin, Hsiu, and Chuang]{yang2018ssr}
Tsun-Yi Yang, Yi-Hsuan Huang, Yen-Yu Lin, Pi-Cheng Hsiu, and Yung-Yu Chuang.
\newblock Ssr-net: A compact soft stagewise regression network for age
  estimation.
\newblock In \emph{IJCAI}, volume~5, pp.\ ~7, 2018.

\bibitem[Zeng et~al.(2018)Zeng, Xu, Lan, Guo, and Cheng]{zeng2018multi}
Wei Zeng, Jun Xu, Yanyan Lan, Jiafeng Guo, and Xueqi Cheng.
\newblock Multi page search with reinforcement learning to rank.
\newblock In \emph{Proceedings of the 2018 ACM SIGIR International Conference
  on Theory of Information Retrieval}, pp.\  175--178, 2018.

\bibitem[Zhang et~al.(2016)Zhang, Zhang, Li, and Qiao]{zhang2016joint}
Kaipeng Zhang, Zhanpeng Zhang, Zhifeng Li, and Yu~Qiao.
\newblock Joint face detection and alignment using multitask cascaded
  convolutional networks.
\newblock \emph{IEEE Signal Processing Letters}, 23\penalty0 (10):\penalty0
  1499--1503, 2016.

\bibitem[Zhang et~al.(2017)Zhang, Song, and Qi]{zhifei2017cvpr}
Zhifei Zhang, Yang Song, and Hairong Qi.
\newblock Age progression/regression by conditional adversarial autoencoder.
\newblock In \emph{IEEE Conference on Computer Vision and Pattern Recognition
  (CVPR)}. IEEE, 2017.

\bibitem[Zhao et~al.(2018)Zhao, Wang, Yatskar, Ordonez, and
  Chang]{zhao2018gender}
Jieyu Zhao, Tianlu Wang, Mark Yatskar, Vicente Ordonez, and Kai-Wei Chang.
\newblock Gender bias in coreference resolution: Evaluation and debiasing
  methods.
\newblock \emph{arXiv preprint arXiv:1804.06876}, 2018.

\bibitem[Zhao(2004)]{zhao2004jump}
Lisa Zhao.
\newblock Jump higher: Analyzing web-site rank in google.
\newblock \emph{Information technology and libraries}, 23\penalty0
  (3):\penalty0 108, 2004.

\bibitem[Zhou \& Agichtein(2020)Zhou and Agichtein]{zhou2020rlirank}
Jianghong Zhou and Eugene Agichtein.
\newblock Rlirank: Learning to rank with reinforcement learning for dynamic
  search.
\newblock In \emph{Proceedings of The Web Conference 2020}, pp.\  2842--2848,
  2020.

\bibitem[Zhou et~al.(2021)Zhou, Zahiri, Hughes, Jadda, Kallumadi, and
  Agichtein]{zhou2021biased}
Jianghong Zhou, Sayyed~M Zahiri, Simon Hughes, Khalifeh~Al Jadda, Surya
  Kallumadi, and Eugene Agichtein.
\newblock De-biased modelling of search click behavior with reinforcement
  learning.
\newblock \emph{arXiv preprint arXiv:2105.10072}, 2021.

\bibitem[Zickuhr et~al.(2012)Zickuhr, Rainie, Purcell, Madden, and
  Brenner]{zickuhr2012libraries}
Kathryn Zickuhr, Lee Rainie, Kristen Purcell, Mary Madden, and Joanna Brenner.
\newblock Libraries, patrons, and e-books.
\newblock \emph{Pew Internet \& American Life Project}, 2012.

\end{thebibliography}
\bibliographystyle{iclr2021_conference}

%\appendix
%\section{Appendix}
%You may include other additional sections here.
\end{document}